\documentclass[12pt,onecolumn,twoside]{IEEEtran}
\usepackage{color}
\usepackage{amssymb}
\usepackage{cite}
\usepackage[cmex10]{amsmath}
\usepackage{algorithmic}
\usepackage{array} 
\usepackage{mathrsfs}
\usepackage{graphicx}
\usepackage{latexsym}
\usepackage{amscd}
\usepackage{amsfonts}
\usepackage{amsmath,amscd,amssymb,verbatim}
\usepackage{graphics}
\usepackage{amsthm}
\usepackage[T1]{fontenc}
\usepackage[utf8]{inputenc}
\usepackage{authblk}
\usepackage{setspace}

\title{On the Blocklength-limited Performance of Relaying under Quasi-Static Rayleigh Channels}

\author[$\dag$]{Yulin Hu, \emph{Student Member, IEEE}, Anke Schmeink, \emph{Member, IEEE}  and James Gross, \emph{Member, IEEE}}

\begin{document}
\doublespacing
%

\title{Blocklength-Limited Performance of Relaying under Quasi-Static Rayleigh Channels}




\maketitle

\begin{abstract}
In this paper, the blocklength-limited performance of a relaying system is studied, where  channels are assumed to experience  quasi-static Rayleigh fading while at the same time only the average channel state information (CSI) is available at the source. 
Both the physical-layer performance (blocklength-limited throughput) and the { {link}}-layer performance (effective capacity) of the relaying system are investigated.  
We propose a simple system operation by introducing a factor based on which we weight the average CSI and let the source determine the coding rate accordingly.
In particular, we prove that both the { {blocklength-limited throughput}} and the effective capacity  are quasi-concave in the weight factor. 
Through numerical investigations, we show the appropriateness of our theoretical model. { {In addition, we observe that relaying is more efficient than direct transmission. Moreover,  this performance advantage of relaying under the average CSI scenario is more significant than under the perfect CSI scenario. Finally, the speed of convergence (between the blocklength-limited performance and the performance in the Shannon capacity regime) in relaying system is faster in comparison to the direct transmission under both the average CSI scenario and the perfect CSI scenario.}}
\end{abstract}


\section{Introduction}
\label{sec:Introduction}

In wireless communications, relaying~\cite{Laneman_2004,  Karmakar_2011, Yuzhen_2014} is 
well known as an efficient way to mitigate wireless fading by exploiting spatial diversity.  
Specifically, two-hop decode-and-forward (DF) relaying protocols significantly improve the capacity
and quality of service~\cite{Wendong_2011, Yulin_2011,My_WD2012_effective_capacity,Bhatnagar_2013}.  
{However, all the above studies of the advantages of relaying are under the ideal assumption of communicating arbitrarily reliably at Shannon's channel capacity, i.e.,  coding is assumed to be performed using a block with an infinite length.}%

{{In the finite blocklength regime, especially when the blocklength is short, the error probability of the communication becomes no longer negligible.}} Recently, an accurate approximation of achievable coding rate is  identified in [8,~Thm.~54] for a single-hop transmission system while taking the error probability into account.  
In~\cite{Verdu_2010} the authors show that the performance loss due to a finite blocklength is considerable and becomes more significant when the blocklength is relatively short.  {{Moreover,   this fundamental study regarding AWGN channels has been extended to Gilbert-Elliott Channels~\cite{Polyanskiy_2011}, quasi-static fading channels~\cite{Yang_2014},~\cite{Gursoy_2013}, quasi-static fading channels with retransmissions~\cite{Peng_2011,Makki_2014} as well as  spectrum sharing networks~\cite{Makki_2015}.  However, all these works focus on single-hop non-relaying systems while the study of the blocklengh-limited performance of relaying is missing.}}

In a two-hop relaying network, relaying exploits spatial diversity but at the same time halves the blocklength of the transmission (if equal time division is considered). 
As has been shown in~\cite{Verdu_2010}  that the performance loss due to a finite blocklength is considerable and becomes more significant when the blocklength is relatively short,  the relaying performance  {in}  the finite blocklength regime becomes interesting.
In our recent work~\cite{Hu_2015}, we address  in general analytical performance
models for relaying with finite blocklengths. We investigate the blocklength-limited throughput (BL-throughput) of relaying, {{where the BL-throughput is 
defined by  the average of correctly decoded bits at the destination per channel use.}
{{{We observe by simulations in~\cite{Hu_2015}  that the performance loss (due to a finite blocklength) of relaying is much smaller than expected, while the performance loss of direct transmission is larger. This observation shows the performance advantage of relaying  {in}  the  finite blocklength regime (in comparison to direct transmission).}}}  
    {{{We further show the reason of this performance advantage in~\cite{Hu_letter_2015} that  relaying has a higher SNR at each hop (in comparison to direct transmission) which makes it set the coding rate more aggressively.}}}

Our previous work in either~\cite{Hu_2015} or~\cite{Hu_letter_2015} is under static channels and with an assumption that {{{the source has perfect channel state information (CSI) of all the links}}}.   
In this paper, we generalize the work in~\cite{Hu_2015} and~\cite{Hu_letter_2015}
 into a scenario with quasi-static Rayleigh channels. 
Under the quasi-static fading model, channels vary from one transmission period to the next. Unlike the static channel model, it may be too optimistic  in practice to have instantaneously perfect CSI of all the fading channels (of the relaying system) {{{at the source}}}. However, if the source does not have perfect CSI, e.g., only the average CSI is available at the source, it is not able to determine an appropriate coding rate to fit the instantaneous channel. 
Thus, the  analysis and improvement of the blocklength-limited performance of such a relaying system (under quasi-static fading channels but  only with average CSI) becomes interesting and also challenging. To the best of our knowledge, these issues have not been studied in detail so far.

Under the described relaying system (with quasi-static fading channels and average CSI), 
 we investigate both the physical-layer performance, e.g., BL-throughput, and the  {{link}-layer performance, e.g., effective capacity\footnote{{{The effective capacity  is a famous performance model that accounts for transmission and queuing effects~\cite{Dapeng_2003}. It characterizes the (maximum) arrival rate of a flow to a queuing system and
relates the stochastic characterization of the service of the queuing system to a queue-length or delay constraints
of the flow. It has been widely applied to the analysis of wireless systems~\cite{Ren_2009,Harsini_2012,Gursoy_2013}. However, no effective capacity analysis
of relaying under finite blocklength model has been published so far to the best of our knowledge.}}}.
We propose a simple system operation by introducing a factor based on which we weight the average CSI and let the source determine the coding rate accordingly.
In particular, we prove that both the  BL-throughput and the effective capacity  are quasi-concave in the weight factor. 
By simulations, we show the  appropriateness of our theoretical model. In addition, we show that  the BL-throughput of relaying is slightly increasing in the blocklength while the effective capacity is significantly decreasing in the blocklength. 
Moreover, we 
observe  the performance advantage of relaying  {in}  the  finite blocklength regime:  Under the condition of having similar Shannon capacity performance,  relaying outperforms direct transmission in the finite blocklength regime.
More importantly, this performance advantage  under the average CSI scenario is more significant than under the perfect CSI scenario.  Finally, we find that the performance loss due to a finite blocklength (the gap between BL-throughput and Shannon/outage capacity) is negligible under the average CSI scenario in comparison to under the perfect CSI scenario.

The rest of the paper is organized as follows. Section~\ref{sec:Preliminaries} describes the system model and briefly introduces the background theory regarding the finite blocklength regime. 
In Section~\ref{sec:Physical_layer_P}, we consider the physical-layer performance and derive the BL-throughput of the relaying system. Subsequently, in  Section~\ref{sec:Higher_layer_P} the link-layer performance of the relaying system is studied, where the distribution of the service process increment and the maximum sustainable date rate are investigated.  
Section~\ref{sec:simulation} presents our simulation results. 
Finally, we conclude our work in Section~\ref{sec:Conclusion}.


\section{Preliminaries}
\label{sec:Preliminaries}

\subsection{System Model}
\label{sec:System_Model}

We consider a simple relaying scenario with a source, a destination and a decode-and-forward (DF) relay as schematically
shown in Figure~\ref{system-instr}. 
\begin{figure}[!h]
\begin{center}
\includegraphics[width=0.55\linewidth, trim=10 20 10 20]{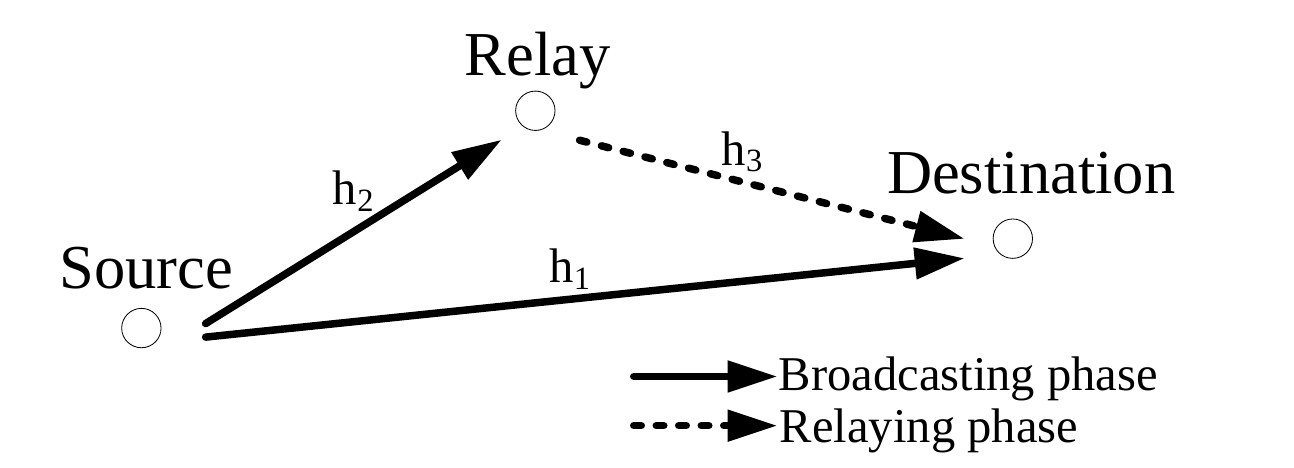}
\end{center}
\caption{Example of the considered single relay system scenario.}
\label{system-instr}
\end{figure}
The links between the above transceivers are referred to as the direct link  (from the
source to the destination), the backhaul link  (from the source to the relay) and the relaying link  (from
the relay to the destination). In general, we assume the direct link to be much weaker than the backhaul
link as well as the relaying link. 
The entire system operates in a slotted fashion  where time is divided into transmission periods of length $2m$ (symbols). Each transmission period contains two frames (each frame with length $m$),  which corresponds to the two hops of relaying.  The blocklength of the coding over the channel in each frame is as long as the frame length $m$. 

During a transmission period $i$, first a broadcasting frame is employed, followed by a relaying frame. In the broadcasting frame, the source transmits data to the relay and the destination. 
The received signals at the destination and the relay  in the broadcasting frame of transmission period $i$ are given by:
\begin{equation}
{\bf{y}}_{1,i}  = p_{\rm{tx}} {{{h}}_{1,i} } {\bf{x}}_i +{\bf{n}}_{1,i},
\end{equation}
\begin{equation}
{\bf{y}}_{2,i} =  p_{\rm{tx}} {h_{2,i} }  {\bf{x}}_i + {\bf{n}}_{2,i}.
\end{equation}
Next, if the data is decoded correctly and forwarded by the relay, the received signal at the destination  in the relaying frame  of transmission period $i$ is given by:
\begin{equation}
 {\bf{y}}_{3,i}  =   p_{\rm{tx}} {{h}}_{3,i}   {\bf{x}}_i  + {\bf{n}}_{3,i}.
\end{equation} 
The transmitted signal $\bf{x}$ and received signals ${\bf{y}}_{1,i}$, ${\bf{y}}_{2,i}$ and~${\bf{y}}_{3,i}$ are complex $m$-dimensional vectors.   Besides, the transmit power at either the relay or the source is denoted by $p_{\rm{tx}}$.   In addition, the noise vectors of these links in transmission period $i$ are denoted by ${\bf{n}}_{1,i}$, ${\bf{n}}_{2,i}$ and~${\bf{n}}_{3,i}$,  which are independent and identically distributed (i.i.d.) complex Gaussian vectors:   ${\bf{n}}$ $\! \sim \! {\mathcal{{N} }}\left( {0,{\sigma ^2}{{\bf{I}}_m}} \right)$,  $ {\bf{n}} \in \{{\bf{n}}_{1,i}, {\bf{n}}_{2,i}, {\bf{n}}_{3,i}\}$, where ${\bf{I}}_m$ denotes an $m \times m$ identity matrix.    
Moreover, ${{h}}_{1,i} $, $h_{2,i}$ and ${{h}}_{3,i}$ are the channels (scalars) of the direct link, backhaul link and relaying link during transmission period $i$, respectively. 
In this work, we consider quasi-static Rayleigh fading channels  where the channels remain constant within each transmission period (includes two hops/frames) and vary independently from one period to the next. 
{{{Hence, the instantaneous channel gain of each link  has two components, i.e., the average channel gain and the random fading. }}} 
On the one hand, {{{we denote  average channel gains (e.g., due to the path loss)  of these three links over the transmission periods by  $|{{\bar h}}_{2}|^2$, $|{{\bar h}}_{2}|^2$ and $|{{\bar h}}_{3}|^2$.}}}
On the other hand, we assume these channels to experience Rayleigh fading where the envelope $\sqrt z $ of the channel fading (response) is Rayleigh distributed with the probability density function (PDF) ${f}\left( {\sqrt z } \right) = 2 \frac{{\sqrt z }}{{{\sigma ^2}}}{e^{ - z/{\sigma ^2}}}$.
Typically, the channel fading is modeled as having unit power gain. This requires that $\sigma ^2 = 1$. Hence, 
the  PDF of  $z$, the gain due to Rayleigh fading, is given by the exponential distribution: $f( z ) = e^{ - z}$. {{{We denote by $z_{j,i}$ ($j = 1,2,3$)  the gains due to Rayleigh fading}}} (during transmission period $i$) of the direct link, the backhaul link and the relaying link.  
{{{Hence, we have $|h_{j,i}|^2 = {z_{j,i}}|{{\bar h}}_{j}|^2, j = 1,2,3$.}  Moreover, these channel fading gains at different links during the same transmission period are assumed to be independent and identically distributed (i.i.d).

Finally, the destination is assumed to apply maximal ratio combining (MRC) 
 where the combined channel gain is given by  $|{{ h}}_{1,i}|^2  + |{{ h}}_{3,i}|^2$.
Thus, the received signal-to-noise ratio (SNR) at the relay and the received SNR at the destination in  transmission period $i$ under maximum ratio combining are given by ${\gamma_{2,i}} = \frac{{{|{{ h}}_{2,i}|^2}{p_{{\rm{tx}}}}}}{{{\sigma ^2}}}$ and ${\gamma_{{\rm{MRC}},i}} = \frac{{\left( {{|{{ h}}_{1,i}|^2} + {|{{ h}}_{3,i}|^2}} \right){p_{{\rm{tx}}}}}}{{{\sigma ^2}}}$.

\subsection{Blocklength-Limited Performance of Single-Hop Transmission Scenario with Perfect CSI {{{(at the source)}}}}
 For  the real additive white Gaussian noise (AWGN) channel,~[8, Theorem 54] derives an accurate approximation of the
coding rate of a single-hop transmission system. With blocklength $m$, block error probability $\varepsilon$ and SNR $\gamma$, the coding rate (in bits per channel use) is given by:
 $r \approx \frac{1}{2}{\log _2}\left( {1 + \gamma } \right) - \sqrt {\frac{V_{\text{real}}}{m}} {Q^{ - 1}}\left( \varepsilon  \right)$, where~$Q^{ - 1}(\cdot)$ is the inverse Q-function, and as usual, the Q-function is given by $Q\left( w \right) = {\rm{ }}\int_w^\infty  {\frac{1}{{\sqrt {2\pi } }}} e^{ - t^2 /2} dt$. In addition, $V_{\text{real}}$ is the  channel dispersion of a real Gaussian channel which is given by $V_{\text{real}} = \frac{\gamma }{2}\frac{{\gamma  + 2}}{{{{\left( {1 + \gamma } \right)}^2}}}{\left( {{{\log }_2}e} \right)^2}$.

{{Under a quasi-static fading channel model, each channel state  is assumed to be static during a transmission period.  Therefore, in each transmission period  a quasi-static fading channel with fading coefficient $h$ can be viewed as an AWGN channel with channel gain $|h|^2$.}} 
Therefore, the above result with a real AWGN channel can be reasonably extended to a complex quasi-static fading channel model  in~\cite{Yang_2014,Gursoy_2013,Peng_2011,Makki_2014,Makki_2015}: with a  channel gain~$|h|^2$ the coding rate of a transmission period (in bits per channel use) is given by: 
\begin{equation}
\label{eq:single_link_data_rate}
r =  {\rm{R}}({|h|^2},\varepsilon ,m)  \approx {\rm{C}}({|h|^2}) - \sqrt {\frac{{{V_{{\rm{comp}}}}}}{m}} {Q^{ - 1}}\left( \varepsilon  \right)\mathrm{,}
\end{equation}
where ${\rm{C}}\left( |h|^2 \right)$ is the  Shannon capacity function of a complex channel with gain~$|h|^2$: $ {\rm{C}}({|h|^2}) \!=\! {\log _2}\left( {1 + \frac{{{|h|^2}{p_{{\rm{tx}}}}}}{{{\sigma ^2}}}} \right)$. 
{{In addition,  the  channel dispersion of a complex Gaussian channel is  twice the one of a real Gaussian channel: ${V_{{\rm{comp}}}} = 2{V_{{\rm{real}}}} = \gamma \frac{{\gamma  + 2}}{{{{\left( {1 + \gamma } \right)}^2}}}{\left( {{{\log }_2}e} \right)^2} = \left( {1 - \frac{1}{{{{\left( {1 + \gamma } \right)}^2}}}} \right){\left( {{{\log }_2}e} \right)^2} = \left( {1 - {{\rm{2}}^{{\rm{ - 2C}}({|h|^2})}}} \right){\left( {{{\log }_2}e} \right)^2}$.}}

Then,  for a single hop transmission of a transmission period of quasi-static fading channel,  
with blocklength $m$ and coding  rate $r$, 
 the decoding (block) error probability at the receiver is given by:  
\begin{equation}
\label{eq:single_link_error_pro}
\varepsilon  = {\rm{P}}({|h|^2},r,m)  \approx Q\left( {\frac{{{\rm{C}}({|h|^2}) - r}}{{\sqrt {{V_{{\rm{comp}}}}{\rm{/}}m} }}} \right).
\end{equation}

{{Considering the channel fading, the expected/average error probability over channel fading is given by~\cite{Yang_2014}:
\begin{equation}
\label{eq:expectedsingle_link_error}
\mathop {\mathop{\rm E}\nolimits} \limits_z \left[ \varepsilon  \right] = \mathop {\mathop{\rm E}\nolimits} \limits_z \left[ {{\rm{P}}(|h{|^2},r,m)} \right]  \approx \mathop {\mathop{\rm E}\nolimits} \limits_z \left[ {Q\left( {\frac{{{\rm{C}}(|h{|^2}) - r}}{{\sqrt {{V_{{\rm{comp}}}}{\rm{/}}m} }}} \right)} \right],
\end{equation}
where $\mathop {\mathop{\rm E}\nolimits} \limits_z \left[ *  \right]$ is the expectation over the distribution of channel fading gain $z$.}}


In the remainder of the paper,  {{{we investigate the blocklength-limited performance of relaying under quasi-static fading channels by  applying the above approximations.}}} As these approximations have been shown to be very tight for sufficiently large value of $m$~\cite{Verdu_2010,Polyanskiy_2011,Yang_2014}, for simplicity we will assume them to be equal in our analysis and numerical evaluation where we consider sufficiently large value of $m$ at each hop of relaying.


\section{Physical-layer Performance of Relaying with Average CSI}
\label{sec:Physical_layer_P}
With only average CSI, if the source determines the coding rate directly based on it, this  likely results in that $|{{ h}}_{j,i}|^2$, the instantaneous channel gain at transmission period $i$, is lower than  the average which is $|{{\bar h}}_{j}|^2$. And this leads to 
a  significant  error probability.
Therefore, 
we propose the source to choose a relatively lower coding rate which is obtained by  the weighted  average channel gains $\eta |{{\bar h}}_{j}|^2$ ($j=1,2,3$), where  $\eta$  is a weight factor. In addition,  we assume that
 $0 < \eta  \le \hat z$, where $\hat z $ is the median of $z$.
{{ For link $j$, with probability  0.5 value $\eta$ is lower than the channel fading gain $z_j$,  i.e., ${\Pr} \{\eta < z_j\}  \le  0.5$.
Hence,  although the instantaneous channel gain $z_j |{{\bar h}}_{j}|^2$ is still possible  to be lower than the weighted one $\eta |{{\bar h}}_{j}|^2$, the probability of this becomes much lower when  $\eta< \hat z$ and is bounded by  0.5, i.e., ${\Pr} \{z_j |{{\bar h}}_{j}|^2 < \eta |{{\bar h}}_{j}|^2\} = {\Pr} \{\eta < z_j\}  \le   0.5$}}\footnote{The setup (letting $ {\hat z} $  be the upper limit of $\eta$) facilitates the proof of Theorem  1  and Theorem  2.  In addition, in practice it is unreasonable to choose a coding rate which  likely exceeds the Shannon limit with a probability higher than 0.5. Our setup just reduces this probability and bounds it by  0.5.    Hence, this setup is reasonable.
Moreover, we will also show in the simulation that this setup is not impacting the system operation/optimization as the optimal value of  $\eta$ is about 0.2 which is much lower than $\hat z \approx  0.7$ (Rayleigh channels).}.

Recall that we assume MRC to be applied at the destination. Hence, the coding rates at different hops of relaying are required to be the same.
{{This coding rate $r$ is  determined by the source based on the weighted average CSI according to~\eqref{eq:single_link_data_rate}.
 As the overall performance of the considered two-hop relaying system is actually mainly subject to the bottleneck link which is either the backhaul link or the combined link,  the coding rate is  determined by
$r = {\rm{R}}({\eta  \cdot \min \left\{{|{{\bar h}_2}{|^2},|{{\bar h}_{\rm{1}}}{|^2} + |{{\bar h}_{\rm{3}}}{|^2}} \right\}},\varepsilon^\circ,m)$, where $\varepsilon^\circ$ is a constant error probability and has a value of practical interest.}}
 According to~\eqref{eq:single_link_data_rate},  $r$ is strictly increasing in the weight factor~$\eta$. In other words, a big $\eta$ means a high expectation on the channel quality and results in a high coding rate. 

Once the coding rate $r$ is determined, it will not be changed during transmission periods. In other words, from one transmission period to the next the coding rate is fixed  and therefore  error probabilities of different links  of relaying vary along with the channel fading.    
{{Regarding the overall error of relaying, in this work we treat the decoding error  at the destination based on the combined channel gain as the overall error. Although it is theoretically possible that an error occurs in the two-hop  relaying though  the direct transmission (in the broadcasting phase) is correct,  the probability of this is negligible as on the one hand we mainly consider  error probabilities of practical interest (which means  that the overall error probability of relaying is not significant) and on the other hand we assume the direct link to be much weaker than the backhaul
link as well as the relaying link.} 
Therefore, the overall error probability of relaying during transmission period~$i$ is given by:
\begin{equation}
\label{eq:Overall_error_pro}
{\varepsilon _{{\rm{R}},i}} = \varepsilon _{2,i} + \left( {1 - \varepsilon _{2,i}} \right) \cdot \varepsilon _{{\rm{MRC}},i} \mathrm{,}
\end{equation}
where $ \varepsilon _{{\rm{MRC}},i}= {\mathop{\rm P}} (|{{ h}}_{1,i}|^2 + |{{ h}}_{3,i}|^2,r,m)$ and $ \varepsilon _{2,i} =  {\mathop{\rm P}} (|{{ h}}_{2,i}|^2,r,m)$.

Under the studied two-hop relaying scenario where the coding rate at each hop is $r$, the (source-to-destination) equivalent coding rate is actually $r/2$. Therefore, the expected BL-throughput of relaying during transmission period $i$ (the number of correctly received bits at the destination per channel use) is given by:
\begin{equation}
\label{eq:Overall_error_pro}
C_{{\rm{BL}},i}= r{(1-\varepsilon _{{\rm{R}},i})} /2  \mathrm{.}
\end{equation}
 
Then, we have the (average) BL-throughput of relaying over time, which is actually the expectation value of $C_{{\rm{BL}},i}$ over the transmission periods: 
\begin{equation}
\label{eq: BL-throughput_over_fading}
{C_{{\rm{BL}}}} = \mathop {\rm{E}}\limits_i \left[ {{C_{{\rm{BL}},i}}} \right] = r(1 - \mathop {\rm{E}}\limits_i \left[ {{\varepsilon _{{\rm{R}},i}}} \right])/2  \mathrm{.}
\end{equation}

Hence, the major challenge for determining ${C_{{\rm{BL}}}}$ is to obtain  the expectation of  the overall error probability over the transmission periods, which is actually equal to the expectation over channel fading. 
Recall that all the channels are independent from each other, hence the expected value of the overall error probability of relaying is further given by: 
\begin{equation}
\label{eq:mean_error_transfer}
\mathop {\rm{E}}\limits_i \left[ {{\varepsilon _{{\rm{R}},i}}} \right]  \!=  \!\! \! \! \mathop {\rm{E}}\limits_{{z_1},{z_2},{z_3}} \!\!\left[ {{\varepsilon _{\rm{R}}}} \right]  \!= \! \mathop {\rm{E}}\limits_{{z_2}}  \!\left[ {{\varepsilon _2}} \right]  + \! ( {1 - \mathop {\rm{E}}\limits_{{z_2}} \left[ {{\varepsilon _2}} \right]}   ) \!\mathop {\rm{E}}\limits_{{z_1},{z_3}} \left[ {{\varepsilon _{{\rm{MRC}}}}} \right] \mathrm{,}
\end{equation}
where $\mathop {\rm{E}}\limits_{{z_2}} \left[ {{\varepsilon _2}} \right]$ and $\mathop {\rm{E}}\limits_{{z_1},{z_3}} \left[ {{\varepsilon _{{\rm{MRC}}}}} \right]$ are the expectation values (over fading) of error probabilities of the backhaul link and the combined link.  {{Based on~\eqref{eq:expectedsingle_link_error}, they are given by:}}
\begin{equation}
\begin{split}
\mathop {\rm{E}}\limits_{{z_2}} \left[ {{\varepsilon _2}} \right] & = \int_0^\infty  {{e^{ - {z_2}}}} {\varepsilon _2}d{z_2}\\
& = {\int_0^\infty  {{\rm{P}}({z_2}|\bar h_2|^2,r,m){e^{ - {z_2}}}dz} _2}\\
& = \frac{1}{{\sqrt {2\pi } }}\int_0^\infty  {\int_{\sqrt m {{w}}({z_2})}^\infty  {{e^{ - \frac{{{t^2} + 2{z_2}}}{2}}}} } dtd{z_2}\mathrm{,}
\end{split}
\label{eq:expected_error_single_2}
\end{equation}
\begin{equation}
\begin{split}
\!\!\mathop {\rm{E}}\limits_{{z_1},\!{z_3}} \! \!\!\left[ {{\varepsilon _{{\rm{MRC}}}}} \right]\! &=\! \int_0^\infty   \! \!\! \!  \int_0^\infty \!   \!\! \! {\varepsilon _{{\rm{MRC}}}} {e^{ - {z_1} \!- {z_3}}}d{z_1}d{z_3}\\
 &=\!  \frac{1}{{\sqrt {2\pi } }}\! \! \int_0^\infty   \! \!\!\!  \int_0^\infty   \! \!\!\!  {\rm{P}}({z_1}|\bar h_1|^2 \!\! +\! {z_3}|\bar h_3|^2\! ,r\! ,m){e^{ \! - {z_1} \! - {z_3}}}d{z_1}d{z_3}\\
& =\!  \frac{1}{{\sqrt {2\pi } }}\! \! \! \int_0^\infty  \! \! \!\!  \int_0^\infty \!  \! \! \! \int_{\sqrt m {{w}}\left( {{z_1},{z_3}} \right)}^\infty \!\!\!\!  {e^{ - \frac{{{t^2} + 2{z_1} + 2{z_3}}}{2}}}dtd{z_1}d{z_3} \mathrm{,}
\end{split} 
\label{eq:expected_error_combine_link}
\end{equation}
where  $w\left( {{z_2}} \right) = \frac{{{\rm{C}}({z_2}|\bar h_2|^2) - r}}{{\sqrt {\frac{1}{m}\left( {1 - {2^{ - 2{\rm{C}}({z_2}|\bar h_2|^2)}}} \right)} {{\log }_2}e}}$ and ${{w}}({z_1},{z_3}) = \frac{{{\rm{C}}({z_1}|\bar h_1|^2 + {z_3}|\bar h_3|^2) - r}}{{\sqrt {\frac{1}{m}\left( {1 - {2^{ - 2 {\rm{C}}({z_1}|\bar h_1|^2 + {z_3}|\bar h_3|^2)}}} \right)} {{\log }_2}e}}$.

So far, we derived the  BL-throughput of relaying under the studied system. We then have the following theorem  regarding the BL-throughput.

\noindent
\textbf {Theorem  1 } \emph {Under a relaying scenario with quasi-static Rayleigh channels where only the average CSI is available at the source, 
the  BL-throughput  is  concave in the coding rate.}

\begin{proof} {See Appendix A.}
\end{proof}

Recall that 
 the coding rate chosen by the source is strictly increasing in the weight factor $\eta$. Combined with   {Theorem~1}, we have an important corollary of Theorem  1:

\noindent
\textbf {Corollary 1} \emph {Consider a relaying scenario 
with quasi-static Rayleigh channels while only the average CSI is available at the source.  If the source determines the coding rate according to the weighted average CSI,  the BL-throughput  is quasi-concave in the weight factor $\eta$.}

\begin{proof} {See Appendix B.}
\end{proof}

Therefore,  only with the average CSI there is a unique optimal value of $\eta$, which maximizes the  BL-throughput of relaying  {in}  the finite blocklength regime.



\

\section{Link-Layer Performance of Relaying with Average CSI}
\label{sec:Higher_layer_P}
In this section, we study the Link-layer performance of relaying based on the \textit{effective capacity}, which is a well-known performance model that accounts for transmission and queuing effects in (wireless) networks~\cite{Dapeng_2003}.  We first briefly review the effective capacity model and extend the  expression of the maximum sustainable data rate into a  blocklength-limited relaying scenario. Subsequently, we derive the MSDR of the studied relaying system  {in}  the finite blocklength regime based on the model.

\subsection{Maximum Sustainable Data Rate}
\label{sec:ER_MSDR}

The effective capacity characterizes the (maximum) arrival rate of a flow to a queuing system and relates the stochastic characterization of the service of the queuing system to the queue-length or delay constraints of the flow. 
{{In}} the finite blocklength regime, decoding errors may occur. If a decoding error occurs in transmission period~$i$, the service process increment (effectively transmitted information)~$s_i$   of the two-frame relaying equals zero. 
On the contrary, if no error occurs at frame $i$, the service process increment equals $s_i = m   \cdot r$, where~$r$ is the coding rate (in bits per channel use) employed over a block of~$m$ symbols in each hop/frame of relaying.  
At period~$i$, the cumulative service process is $S_i = \sum_{n = 0}^{i} s_n$. Assume that the queue is stable as the average service rate is larger than the average arrival rate. Hence, the random queue length $Q_i$ at period $i$ converges to the steady-state random queue length $Q$. To characterize the long-term  statistics~$\mathrm{Pr}\left\{Q\right\}$ of the queue length, the framework of the effective capacity gives us the following upper bound:
\begin{equation}
\mathrm{Pr}\left\{Q > x\right\} \leq K \cdot e^{- \theta \cdot x }  \mathrm{,}
\label{eq:prob_bound}
\end{equation}
where $K$ is the probability that the queue is non-empty and $ \theta$ is the so called QoS exponent.
  Based on~\cite{kumar2004communication}, for a constant rate source with~$r$ bits per two-hop transmission period, the exponent $ \theta$ has to fulfill the following constraint:
\begin{equation}
r < - {{\Lambda \left( { - { \theta}} \right)} \mathord{\left/
 {\vphantom {{\Lambda \left( { - { \theta}} \right)} {{ \theta}}}} \right.
 \kern-\nulldelimiterspace} {{ \theta}}}  \mathrm{.}
\label{eq:qos_requirement}
\end{equation}
$\Lambda\left(\theta\right)$ is the log-moment generating
function of the cumulative service process~$S_i$
defined as:
\begin{equation}
\Lambda\left(\theta\right) = \lim_{i \to \infty} \frac{1}{i} \log \mathrm{E}\left[ e^{\theta \cdot \left(S_i - S_{0}\right)}\right] \! \mathrm{.}
\label{eqn:log_generating_function}
\end{equation}
The ratio $-\Lambda\left(-\theta\right) / \theta$ is called
the effective capacity.
%
%
Denote by $D_i$ the random queuing delay of the head-of-line bit during period $i$. If the  constant arrival rate at the source  is  $r$, with a queue length of $Q = q$, a current delay of the head-of-line bit is given by $D = q/r$. This yields the following approximation for the steady-state delay distribution which is based on Equation~\eqref{eq:prob_bound}:
\begin{equation}
\mathrm{Pr}\left\{D > d\right\} \leq K \cdot e^{- \theta \cdot r \cdot d }  \mathrm{.}
\label{eq:prob_bound_specific}
\end{equation}
%
If the service process $s_i$ can be assumed to be independent and identically distributed (i.i.d.),
a convenient simplification is to obtain the log-moment generating function via the {{{central limit theorem.}}}  
 Then, the effective capacity can be obtained by~\cite{Soret_2009}\footnote{{{{It should be mentioned that~\eqref{eq:single_channel_effective_capacity_central}  holds for i.i.d. $s_i$ but is just an approximation for general (non-i.i.d.) $s_i$}}}}:
\begin{equation}
-\frac{{\Lambda \left( { - \theta } \right)}}{\theta } \!=\!-\! \mathop {\lim }\limits_{i \to \infty } \frac{1}{{i\cdot\theta }}\log {\rm{E}}\left[ {{e^{ - \theta \cdot \sum\nolimits_1^i  {s_i}}}} \right] \!=\! \mathop {\rm{E}}\limits_i \left[ {{s_i}} \right] - \frac{\theta }{2}\mathop {{\rm{Var}}}\limits_i \left[ {{s_i}} \right].
\label{eq:single_channel_effective_capacity_central}
\end{equation}
Therefore, the queuing performance of the system is determined by the mean and the variance
of the random increment of the service process $s_i$. Combining~\eqref{eq:qos_requirement} and~\eqref{eq:prob_bound_specific}, the maximum arrival rate at the source $R_{\rm MS}$ in (bits per transmission period) that can be supported by the random service process  is obtained,  which has been first proposed by~\cite{Soret_2010}:
$ \frac{{\mathop {\rm{E}}\limits_i \left[ {{s_i}} \right]}}{2} + \frac{1}{2}\sqrt {{{( {\mathop {\rm{E}}\limits_i \left[ {{s_i}} \right]} )}^2} + \frac{{2\ln \left( {{P_{\rm{d}}  }} \right)}}{d}\cdot\mathop {{\rm{Var}}}\limits_i \left[ {{s_i}} \right]}$. In the formula, ${\left\{{d,P_{\rm{d}} } \right\}}$ is  QoS requirement pair of the service, where in~\cite{Soret_2010} $d$  is the delay constraint with transmission period as unit and $P_{\rm{d}}$ is the constraint of delay violation probability.

The above maximum source rate  is the so-called maximum sustainable data rate (MSDR) and in this work we refer to MSDR as the metric of link-layer performance. 
Now, we extend the above MSDR into the studied relaying system  {in}  the finite blocklength regime. First, we redefine the unit of the delay constraint $d$ and let $d$ be in symbols.
Recall that in the studied system the length of each (two-hop relaying) transmission period is~$2m$. Therefore,  the delay constraint is $d/2m$  times of a transmission period. Then, the MSDR (bits per channel use)  of the studied two-hop relaying system is given by:
\begin{equation}
 {R_{\rm MS} } \! \approx \! \frac{{\mathop {\rm{E}}\limits_i \left[ {{s_i}} \right]}}{4m} \!+ \! \frac{1}{4m}\sqrt {{{\left( {\mathop {\rm{E}}\limits_i \left[ {{s_i}} \right]} \right)}^2} \!+\! \frac{{4m\ln \left( {{P_{\rm{d}}  }} \right)}}{d}\!\cdot\!\mathop {{\rm{Var}}}\limits_i \left[ {{s_i}} \right]}  \mathrm{.}
\label{eq:rate_function_general}
\end{equation}
Based on~\eqref{eq:rate_function_general}, the major challenge for determining the MSDR ${R_{\rm MS}}$ is to obtain the mean and variance of the service process  increment of the studied relaying system.

\subsection{Mean and Variance of the Service Process Increment}

Recall that the service process increment of a transmission period  is either zero or $rm$. In other words, it is  Bernoulli-distributed. Moreover, the probability of the error event of this Bernoulli distribution is actually the expected overall error probability of relaying which was given in the previous section by~\eqref{eq:mean_error_transfer}.
Based on the characteristic of the Bernoulli distribution, we immediately have the mean and variance of  service process increments:
\begin{equation}
\label{eq:mean_final}
\mathop {\rm{E}}\limits_i \left[ {{s_i}} \right]= rm \cdot (1 - \mathop {\rm{E}}\limits_{{z_1},{z_2},{z_3}} \left[ {{\varepsilon _{\rm{R}}}} \right])  \mathrm{,}
\end{equation}
\begin{equation}
\label{eq:var_final}
\mathop {{\rm{Var}}}\limits_i \left[ {{s_i}} \right] = {r^2}{m^2}\mathop {\rm{E}}\limits_{{z_1},{z_2},{z_3}} \left[ {{\varepsilon _{\rm{R}}}} \right] \cdot (1 -\mathop {\rm{E}}\limits_{{z_1},{z_2},{z_3}} \left[ {{\varepsilon _{\rm{R}}}} \right]) .
\end{equation}

Substituting~\eqref{eq:mean_final} and~\eqref{eq:var_final} into~\eqref{eq:rate_function_general}, we have
\begin{equation}
\label{eq:MSDR_final}
{R_{{\rm{MS}}}} = \frac{{r(1 - \mathop {\rm{E}}\limits_{{z_1}\!,{z_2}\!,{z_3}\!} \left[ {{\varepsilon _{\rm{R}}}} \right])}}{4}  + \frac{r}{4}\sqrt {{{( {1 \!-\!\! \mathop {\rm{E}}\limits_{{z_1}\!,{z_2}\!,{z_3}\!} \left[ {{\varepsilon _{\rm{R}}}} \right]} )}^2} \!\!\!+\! \frac{{4m\ln \left( {{P_{\rm{d}}}} \right)}}{d} \!\!\!\!\mathop {\rm{E}}\limits_{{z_1}\!,{z_2}\!,{z_3}\!} \left[ {{\varepsilon _{\rm{R}}}} \right](1 \!-\!\!\! \mathop {\rm{E}}\limits_{{z_1}\!,{z_2}\!,{z_3}\!} \left[ {{\varepsilon _{\rm{R}}}} \right])} .
\end{equation}

Obviously, the performance of  MSDR $ {R_{\rm MS} }$ is subject to the coding rate $r$, as ${\mathop {\rm{E}}\limits_i \left[ {{\varepsilon _{{\rm{R}},i}}} \right]}$ is also a function of $r$. In fact, the following theorem  holds regarding the relationship between MSDR and the coding rate.

\noindent
\textbf {Theorem  2} \emph {Under a relaying scenario with quasi-static Rayleigh channels where only the average CSI is available at the source,  
 the  MSDR is concave in the coding rate $r$.}

\begin{proof} {See Appendix D.}
\end{proof}

Similar to Theorem  1, Theorem  2 has the following corollary:

\noindent
\textbf {Corollary 2 } \emph {Consider a relaying scenario   
with quasi-static Rayleigh channels while only the average CSI is available at the source.  If the source determines the coding rate according to the weighted average CSI, the MSDR  is quasi-concave in the weight factor $\eta$.}
\begin{proof} {The proof of Corollary 2 (based on  Theorem  2) is similar to the proof of  Corollary~1 (based on  Theorem~1).}
\end{proof}

Hence, the {{link}}-layer performance MSDR also can be optimized by choosing an appropriate~$\eta$. 




\section{NUMERICAL RESULTS AND DISCUSSION}
\label{sec:simulation}

In this section, we first show the appropriateness of our theoretical model. Subsequently, we evaluate the performance of the studied relaying system (with average CSI) in comparison to direct transmission and relaying with perfect CSI. 

For the numerical results, we consider the following parameterization of the system model:
 {{First of all, we consider the cases with blocklength $m \ge 100$  (at each hop of relaying), for
which the approximation is tight enough\footnote{{{The choice of $m \ge 100$ as the minimum
possible length is motivated by [10, Fig. 2] where the relative
difference of the approximate  and the exact achievable rates is
less than 2\% for the cases with $m \ge$ 100.}}}.
In addition, the codewords length of each link is set to be the same as the blocklength.}}
Secondly,  we consider an outdoor urban scenario and the  distances of the backhaul, relaying and direct links are set to 200~$\rm{m}$, 200~$\rm{m}$ and 360~$\rm{m}$. Thirdly, we set the transmit power $p_{\rm{tx}}$  equal to 30~$\rm{dBm}$ and noise power to -90~$\rm{dBm}$,  respectively.  
In addition, we utilize the well-known COST~\cite{Molisch_2011} model for calculating the path-loss while the center frequency is set to equal  2~$\rm{GHz}$. Regarding the channel,  we only consider quasi-static channel model in the simulation.  Hence, the numerical results for all validations and evaluations are based on the average/ergodic performance over the random channel fading. 
Moreover,  as we consider the {{link}}-layer performance, in the simulation the QoS constraints \{delay, delay violation probability\} are set to $\{10^{4}$~symbols, $10^{-2}\}$.  
Finally, to observe the relaying performance we mainly vary the following parameters in the simulation: Blocklength and weight factor~$\eta$.


\subsection{Appropriateness of our theoretical model}

In Figure~\ref{Capacity_codingratewithoutCSI} we show the relationship between the relaying performance and the coding rate. We plot the related (ergodic) Shannon capacity of relaying as a reference. As shown in the figure, the Shannon capacity is not influenced by the coding rate. 
More importantly,  it is shown that the BL-throughput and the MSDR are concave in  the coding rate, which matches Theorem  1 and 2.
During the low coding rate region, both the BL-throughput and the MSDR increase approximately linearly as the coding rate increases.  This is due to the fact that  errors hardly occur with a low coding rate. Hence, the bottleneck of the performance is the coding rate. However, as the coding rate continues increasing, the error probability becomes  more and more significant. As a result, the error probability then becomes the major limit of the system performance.   Therefore, both  the BL-throughput and the MSDR  decrease during the higher coding rate region.
\begin{figure}[!h]
\begin{center}
\includegraphics[width=0.55\linewidth, trim=10 20 10 0 ]{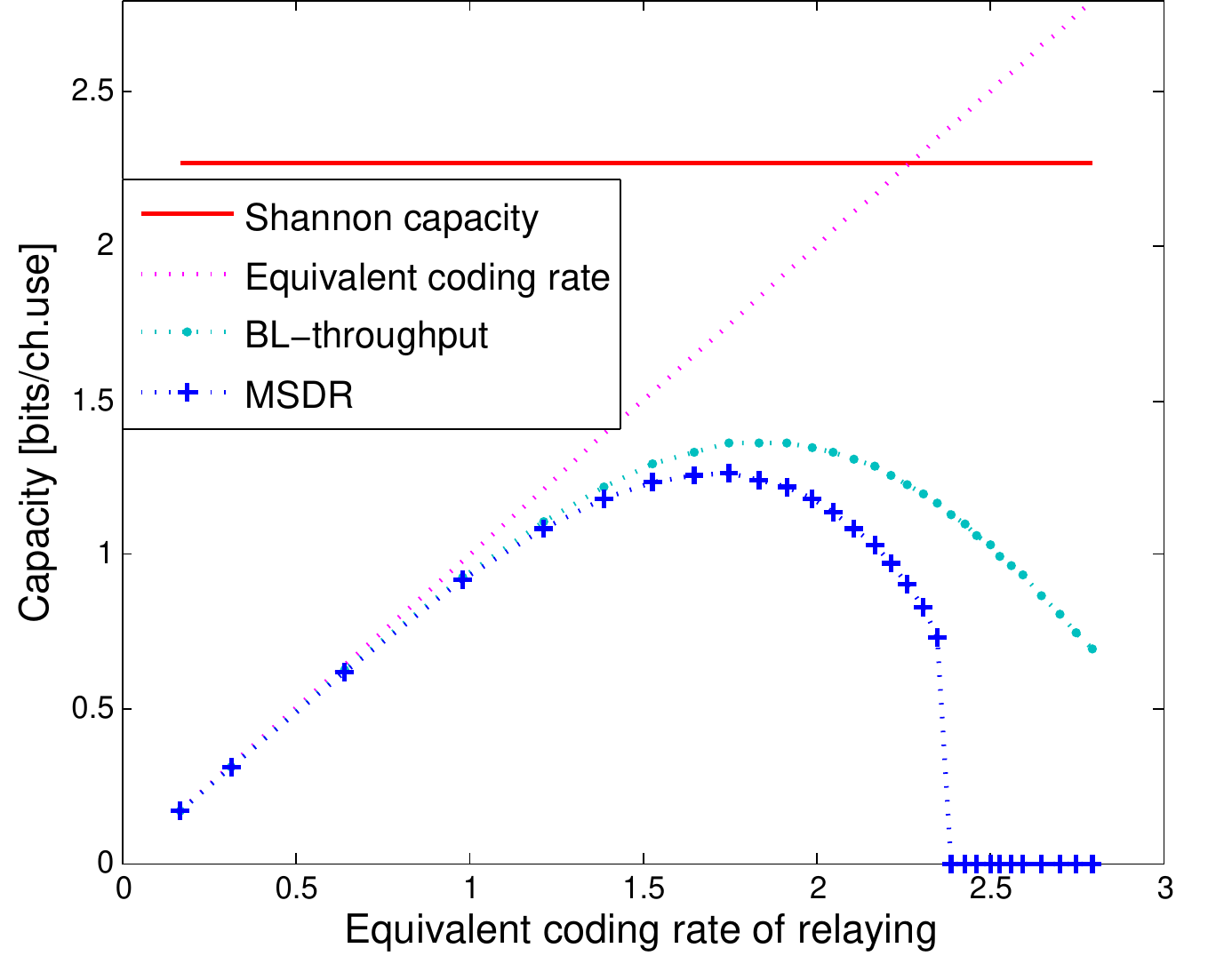}
\end{center}
\caption{The performance of the studied relaying system (average CSI at the source).}
\label{Capacity_codingratewithoutCSI}
\end{figure}

\begin{figure}[!h]
\begin{center}
\includegraphics[width=0.55\linewidth, trim=10 15 10 0 ]{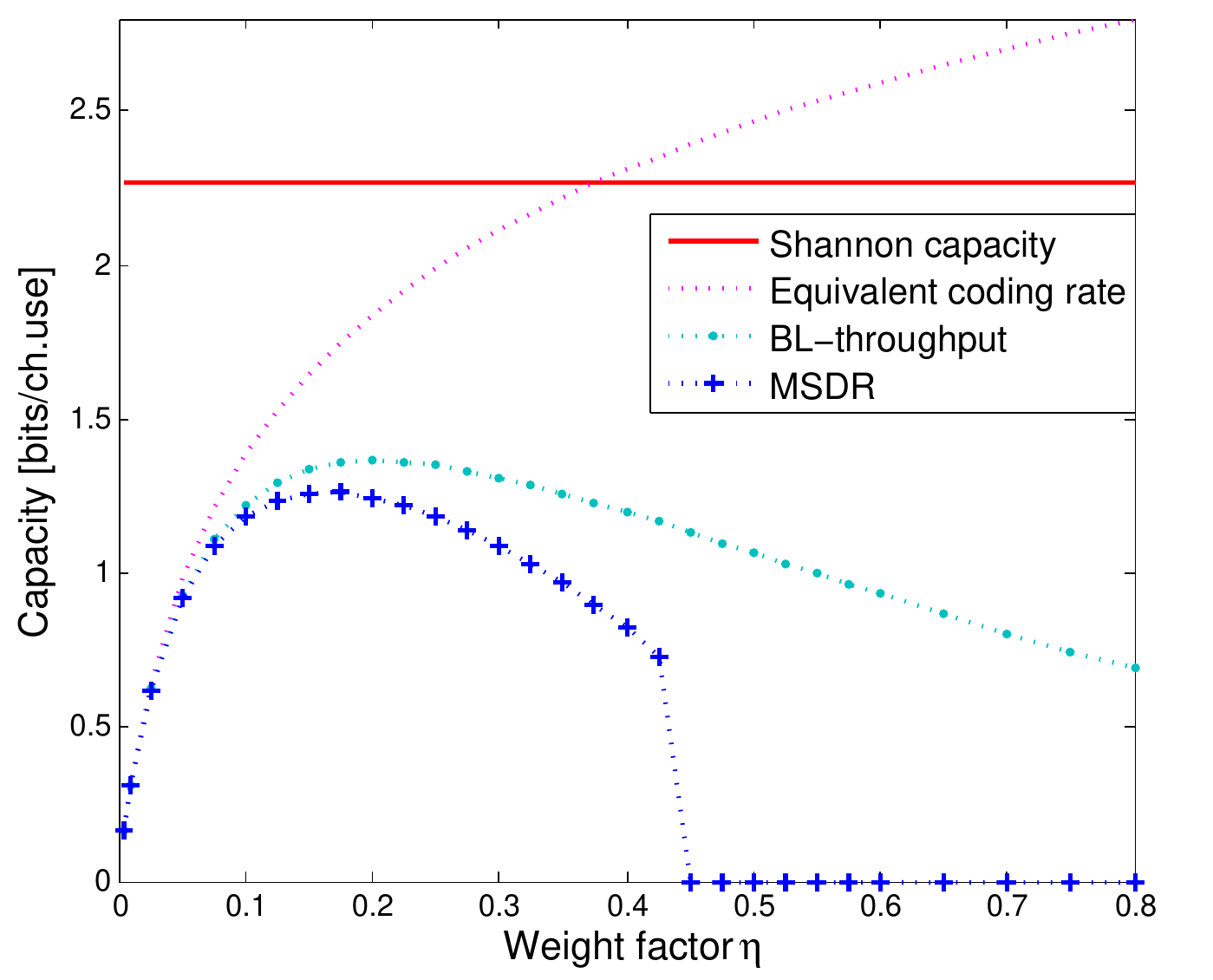}
\end{center}
\caption{The relaying performance versus the weight factor.}
\label{capacityvs_weight}
\end{figure}

Figure~\ref{capacityvs_weight} validates  Corollary 1 and 2 that both  the BL-throughput and the MSDR are quasi-concave in the weight factor $\eta$.
Hence,  the weight factor introduces a good tradeoff between the coding rate and the error probability to the studied relaying system (only with average CSI at the source). Based on this tradeoff, both the physical-layer performance and the {{link}-layer performance of the system can be optimized. Moreover, it  can also be observed from Figure~\ref{capacityvs_weight}  that the optimal values of the weight factor for maximizing the BL-throughput and for maximizing the MSDR are not the same. 
Moreover, we also find in the simulation (not shown here) that the optimal $\eta$ for maximizing the MSDR is subject to the QoS constraints. In particular, the stricter the constraints are the smaller the optimal $\eta$ is.
Hence, the parameterization of  the weight factor $\eta$ for a QoS-support system and a non-QoS-support system should be treated differently. 
These are important guidelines for the design of the studied relaying system.

In Figure~\ref{weightfactorwork}, we provide intuitive parallel sub-figures  to show how the weight factor introduces the tradeoff between the coding rate and the error probability and further influences the BL-throughput and the MSDR.
\begin{figure}[!h]
\begin{center}
\includegraphics[width=0.55\linewidth, trim=10 15 10 20 ]{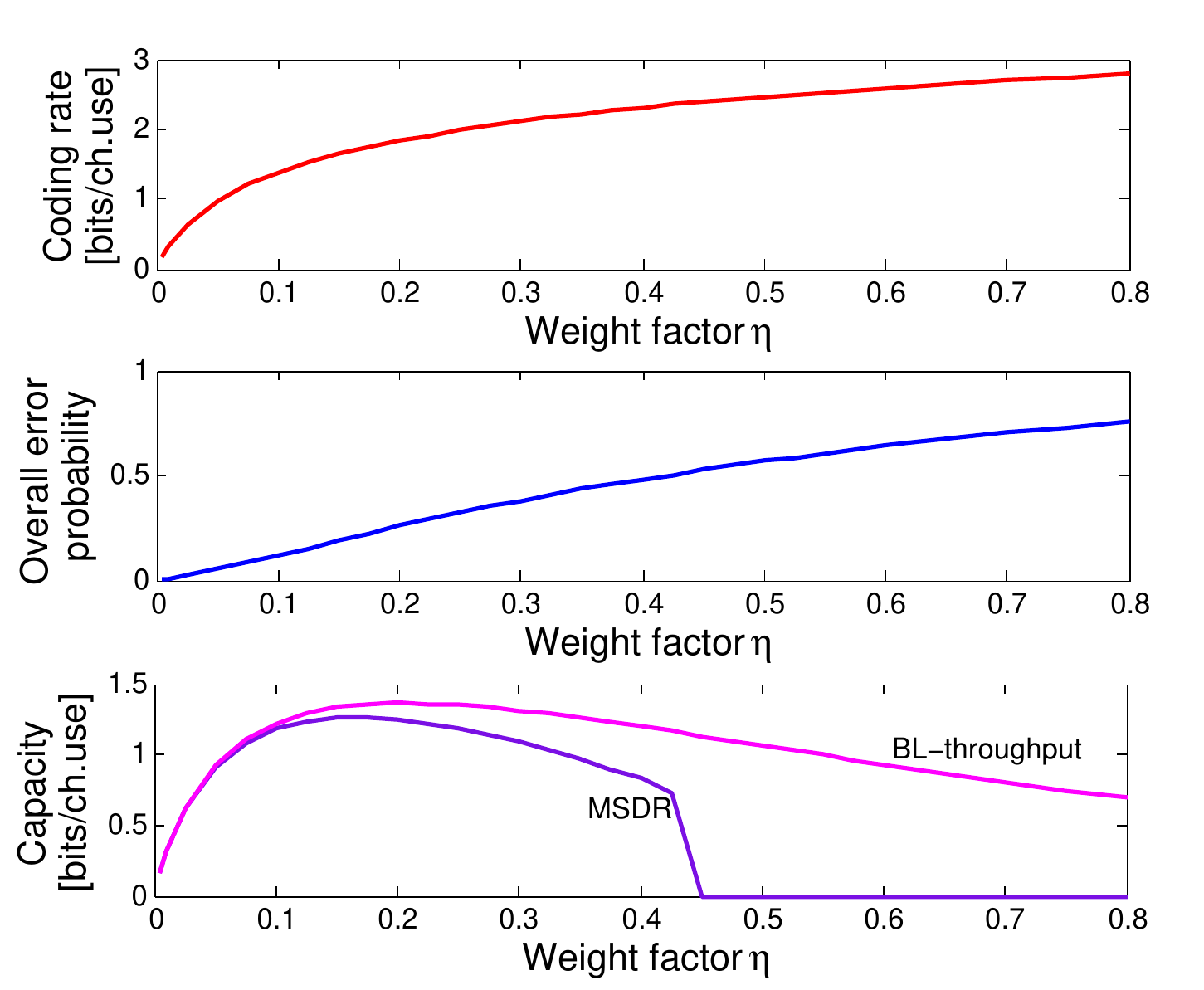}
\end{center}
\caption{How the weight factor introduces tradeoff between error probability and coding rate and further influences the BL-throughput and the MSDR?   In the simulation, the blocklength at each hop of relaying is 500 symbols.}
\label{weightfactorwork}
\end{figure}
As shown in the figure, by increasing the weight factor the coding rate increases as well as the error probability.  In particular, the coding rate increases rapidly at the beginning but slowly later on. At the same time, the error probability increases approximately linearly. 
As a result, the BL-throughput/MSDR first increases and then deceases. In other words, they are quasi-concave in the weight factor  $\eta$.

\subsection{Evaluation}
So far, we have shown the appropriateness of our Theorems and corollaries. 
In the following, we further evaluate the studied relaying system. {{{In the evaluation, the performance under the perfect CSI scenario is considered as a contrast. In particular, a result of the (average) BL-throughput with the perfect CSI is obtained by maximizing the instantaneous BL-throughput\footnote{{{In~[15] we show that the BL-throughput of relaying can be maximized by choosing an appropriate coding rate $r$ (based on the perfect CSI) for AWGN channels. This also holds for maximizing the instantaneous BL-throughput at each transmission period under the  quasi-static channels.}}} for each transmission period based on the perfect CSI.}}}

\subsubsection{Under average CSI scenario: relaying performance vs. blocklength}
We first show the relationship between the performance of the studied relaying system and the blocklength in Figure~\ref{capacity_blocklength}. The figure shows that the BL-throughput of the studied relaying system is slightly increasing in the blocklength while the MSDR is significantly decreasing in the blocklength. The explanation is as follows. 
\begin{figure}[!h]
\begin{center}
\includegraphics[width=0.55\linewidth, trim=10 15 -0 10 ]{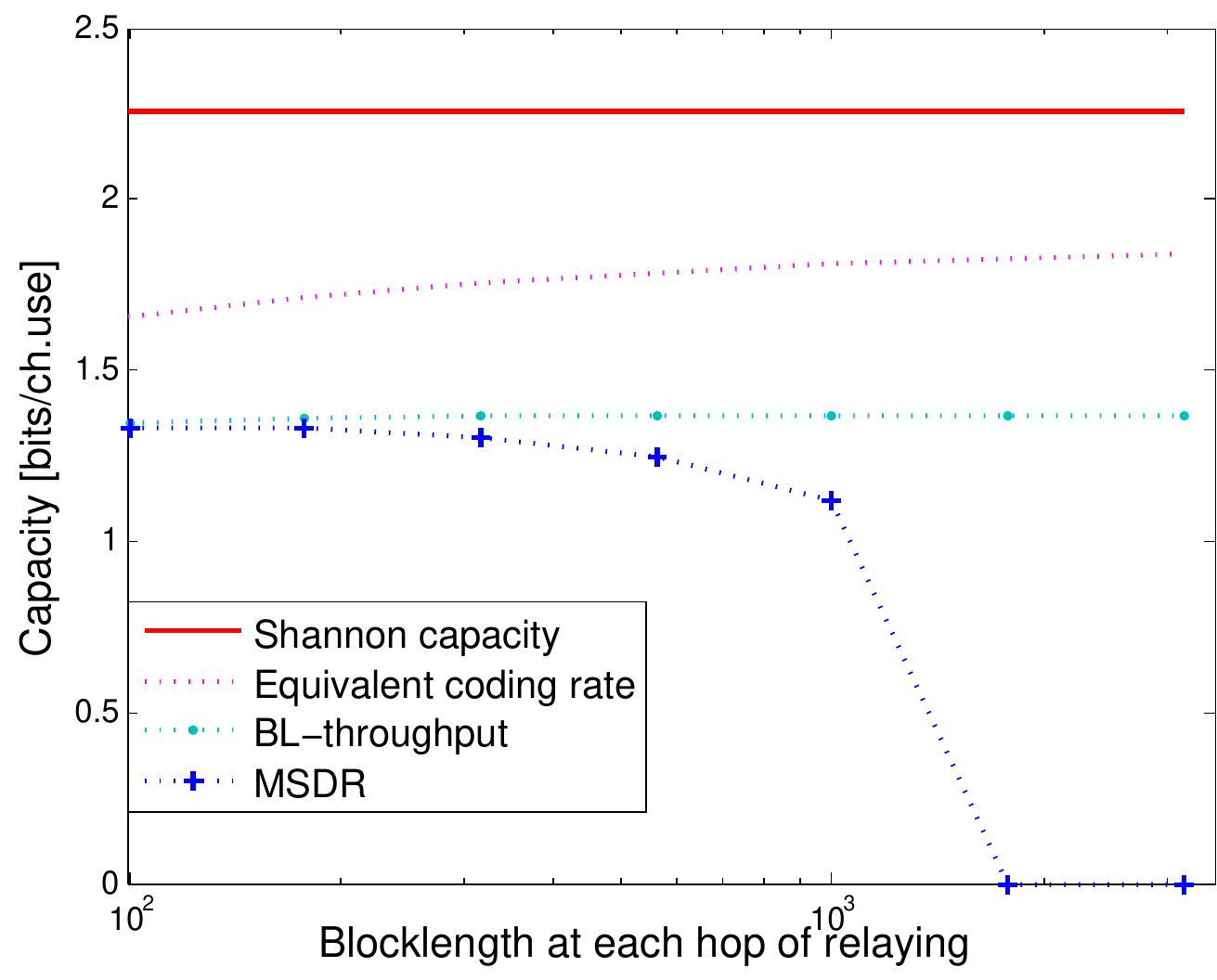}
\end{center}
\caption{The relaying performance versus blocklength.}
\label{capacity_blocklength}
\end{figure}
On one hand, based on~\eqref{eq:single_link_error_pro} and~\eqref{eq:single_link_data_rate} both error probability and coding rate are influenced by the blocklength $m$. In particular, with a fixed coding rate at each link a long blocklength leads to a low error probability of each link. Obviously, this results in a low overall error probability of relaying and therefore a high BL-throughput. This is the reason why the BL-throughput is slightly increasing in the blocklength. On the other hand, a long blocklength means that a single transmission (i.e., two-hop relaying) costs a long time. This reduces the number of allowed retransmission attempts under a given delay constraint. For example, consider a relaying system where the blocklength of each hop of relaying is $500$ symbols (two-hop relaying transmission period is $10^{3}$ symbols). 
To support a service with a delay constraint $d = 10^{4}$ symbols,  the maximal number of transmission attempts (including an initial transmission and retransmissions),  which not violates the delay constraint, is 10  times.  This number would be 5 if the blocklength is doubled. In other words, a long blocklength reduces the flexibility of a QoS-support system due to limiting the number of transmission attempts.     At the same time, the gain from increasing blocklength becomes tiny as the blocklength increases (e.g., even ragarding the physical-layer performance, the BL-throughput is approximately a constant during the long blocklength region).   As a result, the MSDR reduces significantly in the long blocklength region. For an extreme example, the MSDR is zero if the blocklength is longer than the delay constraint. 


\subsubsection{Under average CSI scenario: relaying vs. direct transmission --- observing the performance advantage of relaying}

Under the average CSI scenario, we compare the  performance of the studied relaying system with  direct transmission.
%
To make a fair comparison, we set the coding rate of direct transmission to be equal to the equivalent coding rate of relaying.  In addition, we set the blocklength of direct transmission to be twice  as large as the blocklength at each hop of relaying. Hence, the length of each transmission period of relaying equals that of direct transmission.  
\begin{figure}[!h]
\begin{center}
\includegraphics[width=0.55\linewidth, trim=10 18 10 7 ]{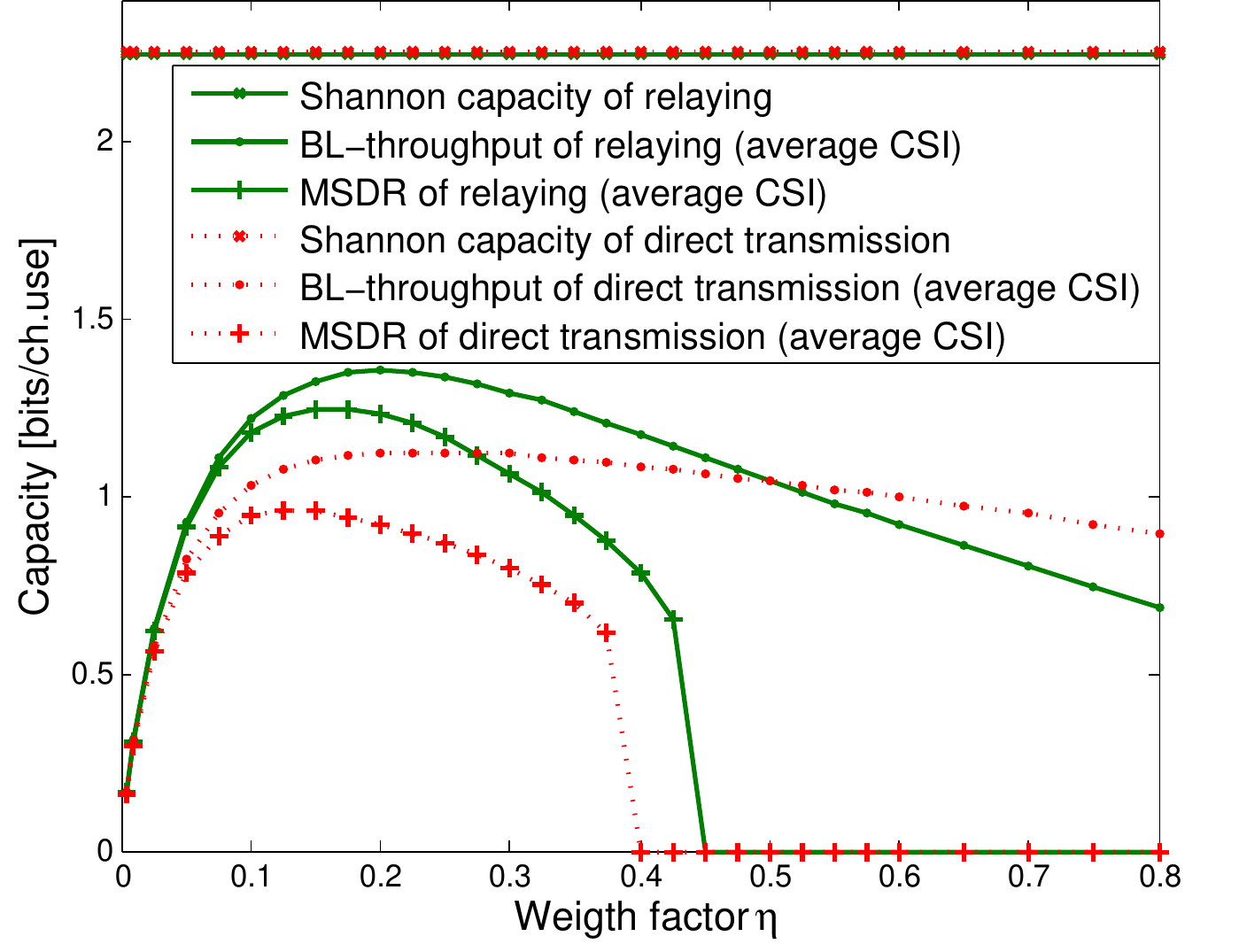}
\end{center}
\caption{The performance comparison between relaying (with average CSI) and direct transmission (with average CSI) while varying the weight factor $\eta$.   In the simulation, the blocklength at each hop of relaying is 500 symbols. }
\label{Relayingvsderct_codingrate}
\end{figure}
\begin{figure}[!h]
\begin{center}
\includegraphics[width=0.55\linewidth, trim=10 18 -15 10 ]{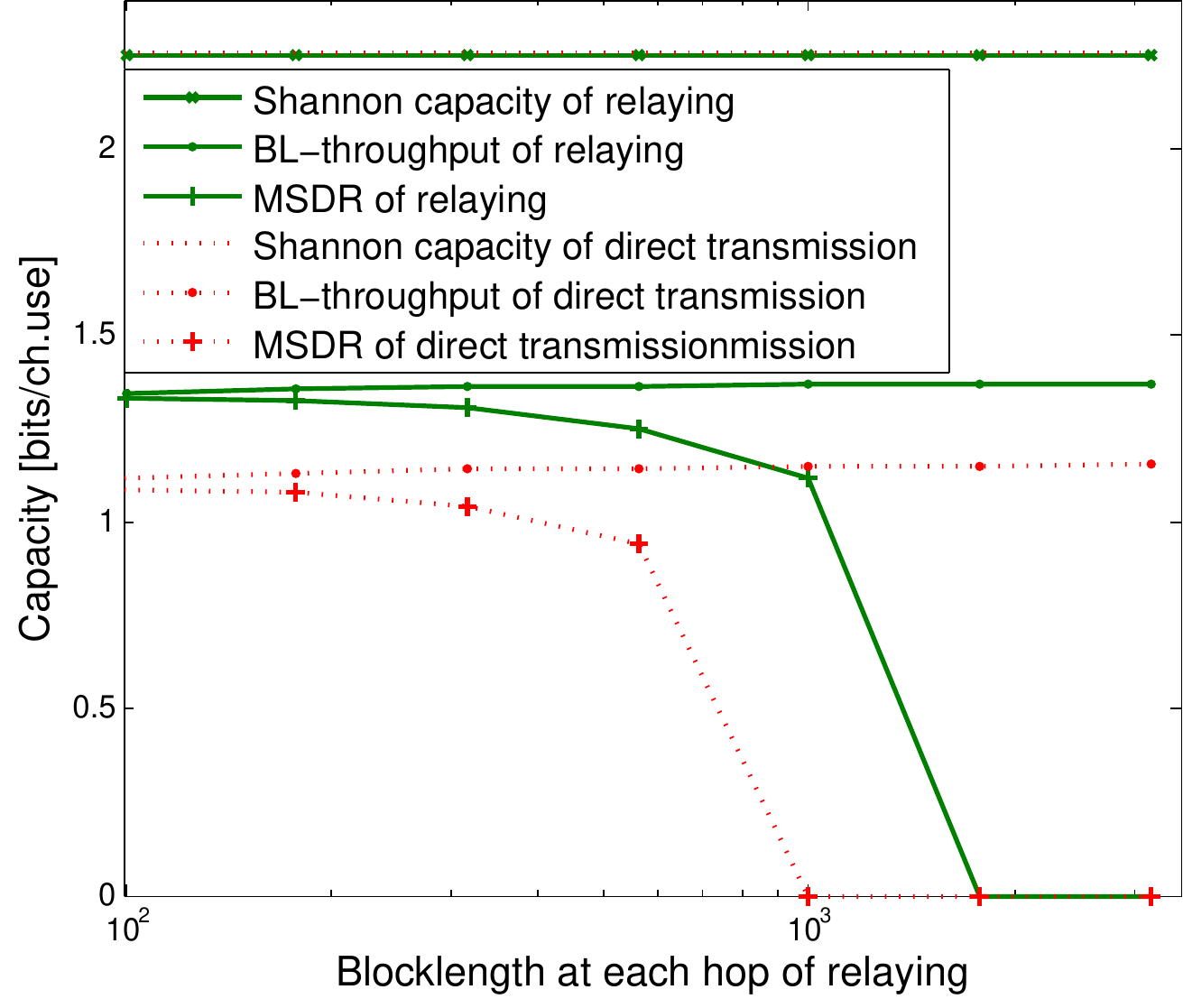}
\end{center}
\caption{The performance comparison between relaying (with average CSI) and direct transmission (with average CSI) under dynamic blocklength ($\eta=0.2$).}
\label{Relayingvsderct_blocklength}
\end{figure}
Under the above setup, we compare the two schemes in Figure~\ref{Relayingvsderct_codingrate} and Figure~\ref{Relayingvsderct_blocklength}
where we vary the coding rate and blocklength, respectively.
Based on these two figures, we observe that {{both the BL-throughput and the MSDR of relaying significantly outperform direct transmission}},  although these two schemes have a similar Shannon capacity.  
In other words, under the average CSI scenario relaying shows a significant advantage (in comparison to the direct transmission)  on the blocklength-limited performance.  

\subsubsection{The performance advantage of relaying: under perfect CSI scenario vs. under average CSI scenario}
The previous subsection shows the performance advantage of relaying  under the average CSI scenario. In this subsection, we compare this performance advantage under the  average CSI scenario with the one under the perfect CSI scenario. 
We show the comparison in  Figure~\ref{averageCSI_vs_perfectCSI_weight} and Figure~\ref{averageCSI_vs_perfectCSI_blocklength} where the weight factor and the blocklength are varied respectively.
\begin{figure}[!h]
\begin{center}
\includegraphics[width=0.55\linewidth, trim=10 12 10 15 ]{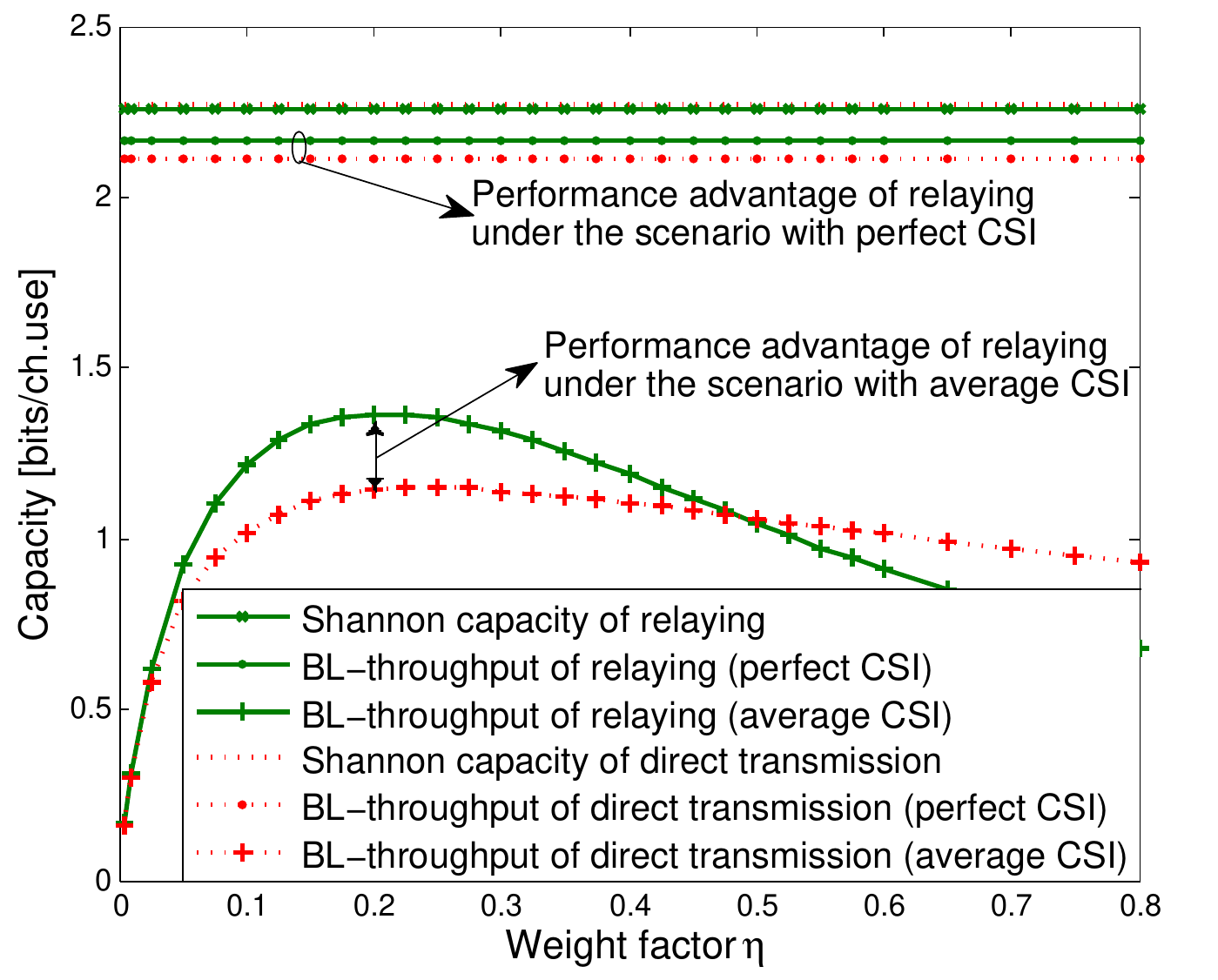}
\end{center}
\caption{The performance comparison between relaying with average CSI and relaying with perfect CSI while varying the weight factor ($\eta=0.2$).}
\label{averageCSI_vs_perfectCSI_weight}
\end{figure}
\begin{figure}[!h]
\begin{center}
\includegraphics[width=0.55\linewidth, trim= 10 19 10 12 ]{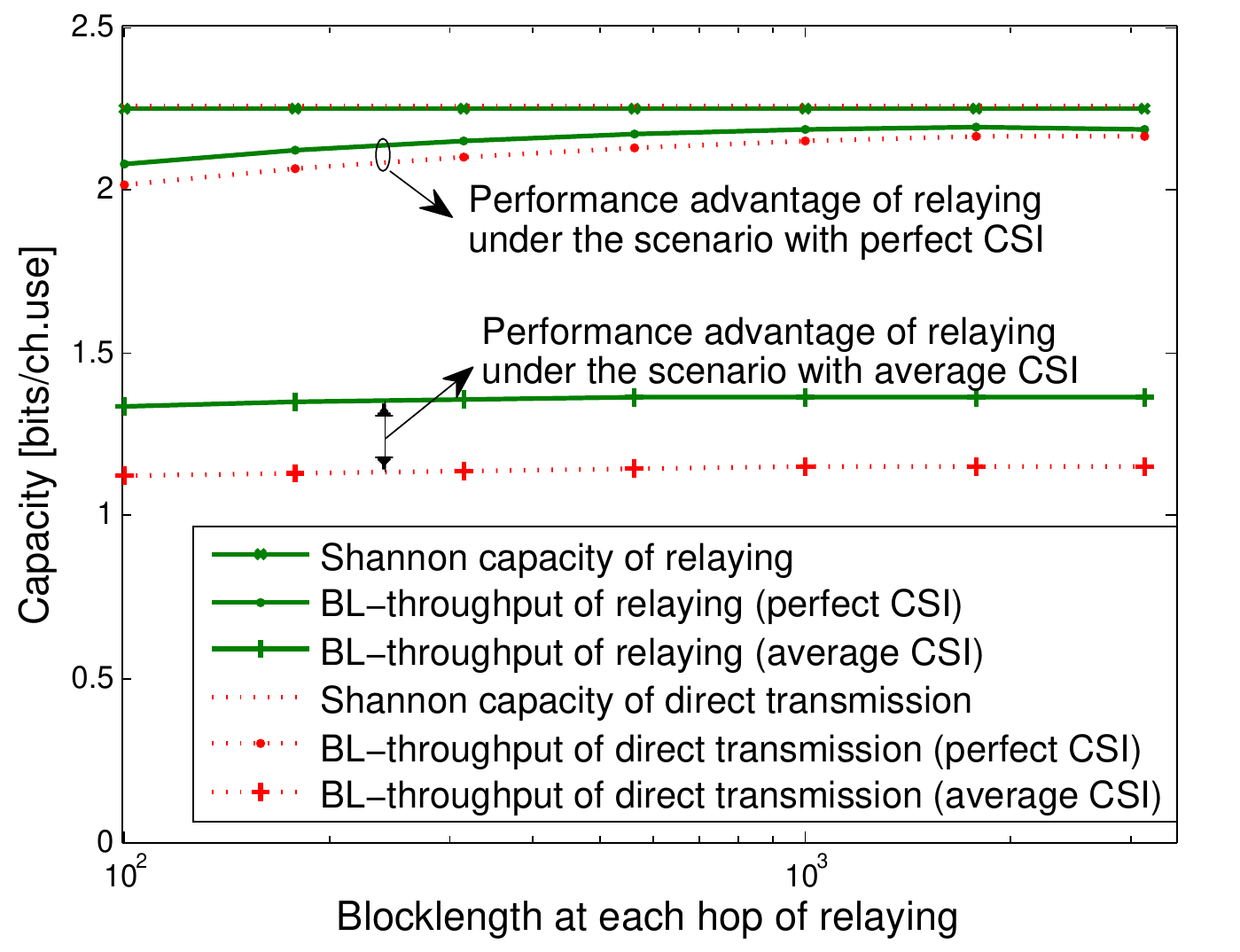}
\end{center}
\caption{The performance comparison between relaying with average CSI and relaying with perfect CSI under dynamic blocklength ($\eta=0.2$).}
\label{averageCSI_vs_perfectCSI_blocklength}
\end{figure}
In general, these figures show that relaying has better blocklength-limited performance than direct transmission under both the perfect CSI scenario and the  average CSI scenario. 
However, we further observe that this performance advantage of relaying under the average CSI scenario is more significant than under the perfect CSI scenario, i.e.,  the performance improvement by relaying {{(comparing relaying with direct transmission)}} under the  average CSI scenario  is significantly higher than the one under the perfect CSI scenario. 
{{{Recall that in~\cite{Hu_2015} we have shown that  with perfect CSI relaying is more beneficial  {in}  the finite blocklength regime in comparison to  {in}  the Shannon capacity regime.
 The observation from Figure~\ref{averageCSI_vs_perfectCSI_weight} and Figure~\ref{averageCSI_vs_perfectCSI_blocklength} further indicates that}}}  in  the finite blocklength regime  relaying is more beneficial for the average CSI scenario in comparison to the  perfect CSI scenario.  This is another important guideline for the design of blocklength-limited systems.

\subsubsection{The performance loss due to a finite blocklength: under perfect CSI scenario vs. under average CSI scenario}

Finally, we compare the performance losses due to a finite blocklength under perfect CSI scenario and under average CSI scenario. 
 {{Under the perfect CSI scenario, the performance loss due to a finite blocklength is actually the performance gap between the Shannon capacity and the BL-throughput (with perfect CSI).
On the other hand, this performance loss under the average CSI scenario is observed by comparing the BL-throughput (with average CSI) with the outage capacity.}} 
 In particular,  the  outage capacity  is a performance metric  {in}  the Shannon capacity regime. 
  It is given by $r \left( {1 - {{\Pr }_{{\rm{out}}}}(r )} \right)$, where~${{{\Pr }_{{\rm{out}}}}(r )}$ is the outage probability.
To calculate the outage capacity, the weighting CSI operation is also considered in the simulation, i.e., the packet size is chosen based on the Shannon capacity  of the weighted CSI. 

\begin{figure}[!h]
\begin{center}
\includegraphics[width=0.55\linewidth, trim=20 35 10 10 ]{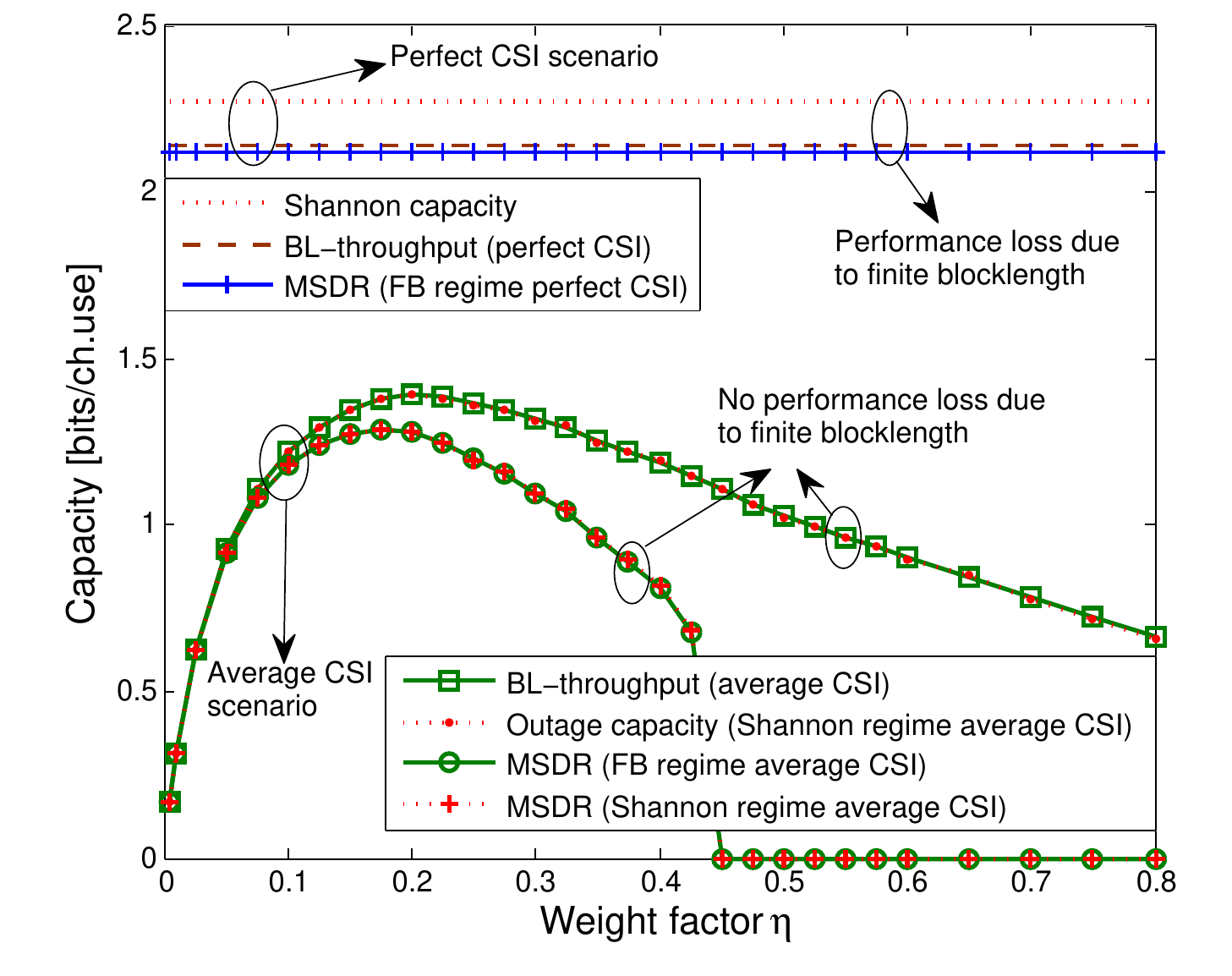}
\end{center}
\caption{The performance comparison between relaying  {in}  the Shannon capacity regime and relaying  {in}  the finite blocklength regime (blocklength at each hop of relaying is 500 symbols).}
\label{fbRelaying_vs_Shannonrelaying}
\end{figure}
Under the above setup, we first show the numerical results of the comparison in Figure~\ref{fbRelaying_vs_Shannonrelaying} where we fix the blocklength and vary the weight factor. 
The figure shows that the  performance loss due to a finite blocklength is considerable under the perfect CSI scenario. 
However, we find that under the average CSI scenario the performance loss due to a finite blocklength is negligible. 
In the simulation, we also observe that (not shown here)  only with average CSI at the source the outage probability ({in}  the Shannon capacity regime) and the average error probability ({in}  the finite blocklength regime) have similar performance. 

The observations from Figure~\ref{fbRelaying_vs_Shannonrelaying} are based on the   setup that the blocklength equals $m=500$ symbols.
Recall that the BL-throughput is limited by the blocklength and the performance  {in}  the Shannon capacity regime is not influenced by the blocklength. Hence, the performance loss due to a finite blocklength should also be subject to the blocklength. 
Then, we further investigate this performance loss in Figure~\ref{fbRelaying_vs_Shannonrelaying_vary_m} where we vary the blocklength.
\begin{figure}[!h]
\begin{center}
\includegraphics[width=0.55\linewidth, trim=10 15 10 10 ]{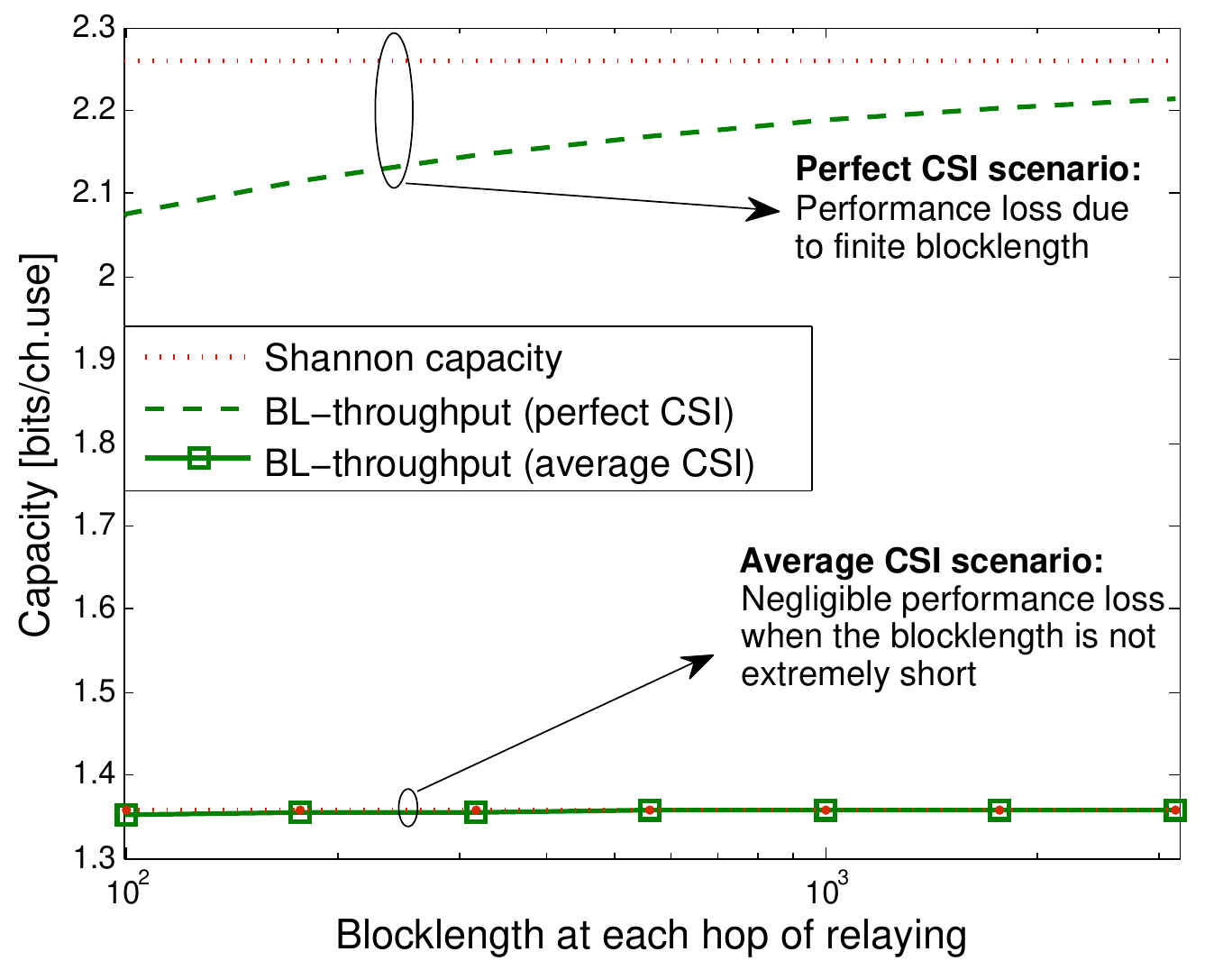}
\end{center}
\caption{The performance comparison between relaying  {in}  the Shannon capacity regime and relaying  {in}  the finite blocklength regime while varying the blocklength. In the simulation, $\eta=0.2$.}
\label{fbRelaying_vs_Shannonrelaying_vary_m}
\end{figure}
Again, we observe  that the performance loss under the average CSI scenario is much smaller than the one under the perfect CSI scenario.  In particular, with a non-extremely short blocklength (e.g., blocklength $m >100$) the performance loss due to a finite blocklength  is negligible under the average CSI scenario.

From Figure~\ref{fbRelaying_vs_Shannonrelaying_vary_m} we observe that the BL-throughput of relaying with average CSI and the outage capacity converge quickly.   It should be mentioned that this is not a unique characteristic of relaying. It is studied and observed in~\cite{Yang_2014} that  the  speed of convergence between the BL-throughput and the outage capacity for a non-relaying direct transmission is also fast. This motivates us to compare the speeds of convergence of these two transmissions schemes. We show the comparison in Figure~\ref{Thelastone}.
\begin{figure}[!h] 
\begin{center}
\includegraphics[width=0.55\linewidth, trim=20 15 0 10 ]{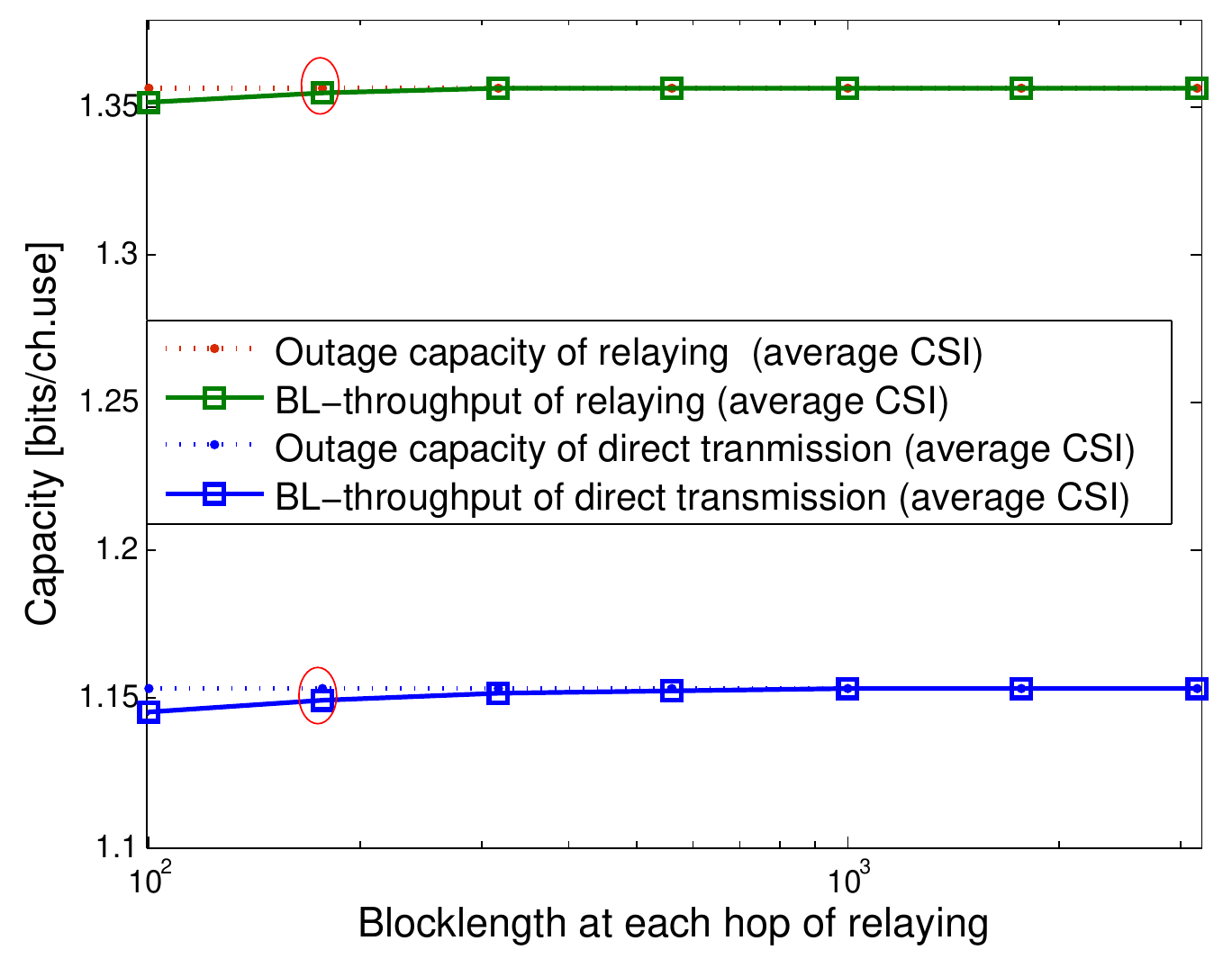}
\end{center}
\caption{The speed of convergence between the BL-throughput with average CSI and  the outage capacity. }
\label{Thelastone}
\end{figure}
From the figure, we first observe that relaying has a slightly faster speed of convergence in comparison to direct transmission. In addition, in general both these two schemes  have quick speeds of convergence between the BL-throughput with average CSI and the outage capacity. This  observation is different from the case with perfect CSI shown in Figure~\ref{averageCSI_vs_perfectCSI_blocklength} where  speeds of the convergence (between the BL-throughput and the Shannon capacity) for both relaying and direct transmission are relatively slow. More interesting,  Figure~\ref{averageCSI_vs_perfectCSI_blocklength} shows that  the speed of convergence  of relaying is also faster than direct transmission under the  perfect CSI scenario. Combining Figure~\ref{averageCSI_vs_perfectCSI_blocklength}   and Figure~\ref{Thelastone}, we conclude that (although halving the blocklength) relaying surprisingly has a  faster  speed of convergence (between the blocklength-limited performance and the performance in the Shannon capacity regime) in comparison to direct transmission under both the average CSI scenario and the perfect CSI scenario. {{{This is actually the performance advantage of relaying  in the finite blocklength regime from a perspective of the speed of convergence (to the Shanon/outage capacity).}}}


\section{Conclusion}
\label{sec:Conclusion}

Under the finite blocklength regime, we investigated the  physical-layer performance (BL-throughput) as well as the {{link}}-layer performance (MSDR) of a relaying system with quasi-static fading channels but only the average CSI (at the source). 
We proposed a simple system operation by introducing a factor based on which we weight the average CSI and let the source determine the coding rate based on the weighted CSI. The BL-throughput  and the MSDR of the studied relaying system are investigated.  
Moreover, we proved that both the  BL-throughput and the MSDR are  concave in the coding rate.
More importantly, it was proved that the  BL-throughput and the MSDR are  quasi-concave in the weight factor. In other words, the performance of the studied system can be easily optimized based on the proposed operation.

We concluded a set of guidelines for the design of efficient relaying systems (under the finite blocklength regime) from numerical analysis.
Firstly, under the average CSI scenario the BL-throughput of relaying is slightly increasing in the blocklength while the effective capacity is significantly decreasing in the blocklength. Hence, determining the blocklength is very important for the design of QoS-support relaying systems. 
Secondly, the weight factor we proposed introduces a good tradeoff between the coding rate and the error probability to the studied relaying system. Based on this tradeoff, both the physical-layer performance and the {{link}}-layer performance of the system can be optimized. Moreover, the optimal values of the weight factor for maximizing the BL-throughput and for maximizing the MSDR are  different.  
Thirdly, under the condition of having similar Shannon capacity performance relaying outperforms direct transmission in the finite blocklength regime. This is actually the performance advantage of relaying in the  finite blocklength regime. 
More importantly,  this performance advantage of relaying under the average CSI scenario is more significant than under the perfect CSI scenario. 
In addition,  the performance loss due to a finite blocklength (i.e., the performance gap between the performance in the Shannon capacity regime and in the finite blocklength regime) is negligible under the average CSI scenario in comparison to the one under the perfect CSI scenario. {{Moreover, the speed of convergence (between the blocklength-limited performance and the performance in the Shannon capacity regime) in relaying system is faster in comparison to the direct transmission under both the average CSI scenario and the perfect CSI scenario.}}

\appendices

\section{Proof of the Theorem  1}

\begin{proof}
Based on Equation~\eqref{eq: BL-throughput_over_fading} we immediately have
\begin{equation}
\frac{{\partial {C_{{\rm{BL}}}}}}{{\partial r}} = \frac{1}{2}(1 - \mathop {\rm{E}}\limits_i \left[ {{\varepsilon _{{\rm{R}},i}}} \right]) - \frac{r}{2}\frac{{\partial \mathop {\rm{E}}\limits_i \left[ {{\varepsilon _{{\rm{R}},i}}} \right]}}{{\partial r}},
\end{equation}
\begin{equation}
\frac{{{\partial ^2}{C_{{\rm{BL}}}}}}{{{\partial ^2}r}} =  - \frac{{\partial \mathop {\rm{E}}\limits_i \left[ {{\varepsilon _{{\rm{R}},i}}} \right]}}{{\partial r}} - \frac{r}{2}\frac{{{\partial ^2}\mathop {\rm{E}}\limits_i \left[ {{\varepsilon _{{\rm{R}},i}}} \right]}}{{{\partial ^2}r}}.
\end{equation}

In the following, we prove  {Theorem  1} by showing $\!\frac{{{\partial ^2}{C_{{\rm{BL}}}}}}{{{\partial ^2}r}}\!\!<\!0$.
 

According to Equation~\eqref{eq:mean_error_transfer}, we have:
\begin{equation}
\label{eq:pr_e_ra_dir}
 \frac{{\partial {\mathop {\rm{E}}\limits_i \left[ {{\varepsilon _{{\rm{R}},i}}} \right]}}}{{\partial r}} = \frac{\partial \mathop {\rm{E}}\limits_{{z_2}} \left[ {{\varepsilon _2}} \right]}{{\partial r}}\left( {1 - {\mathop {\rm{E}}\limits_{{z_1},{z_3}} \left[ {{\varepsilon _{{\rm{MRC}}}}} \right]}} \right) 
 +\frac{\partial \mathop {\rm{E}}\limits_{{z_1},{z_3}} \left[ {{\varepsilon _{{\rm{MRC}}}}} \right]}{{\partial r}}  \left( {1 -{\mathop {\rm{E}}\limits_{{z_2}} \left[ {{\varepsilon _2}} \right]}} \right) \geqslant 0, 
\end{equation}
\begin{equation}
\label{eq:second_de_overall_error_pro}
\begin{split}
 \frac{{\partial ^2 {\mathop {\rm{E}}\limits_i \left[ {{\varepsilon _{{\rm{R}},i}}} \right]}}}{{\partial ^2 r}} &= \frac{{\partial ^2{\mathop {\rm{E}}\limits_{{z_2}} \left[ {{\varepsilon _2}} \right]}}}{{\partial ^2 r}}\left( {1 - {\mathop {\rm{E}}\limits_{{z_1},{z_3}} \left[ {{\varepsilon _{{\rm{MRC}}}}} \right]} } \right) 
- 2\frac{\partial \mathop {\rm{E}}\limits_{{z_2}} \left[ {{\varepsilon _2}} \right]}{{\partial r}}\frac{\partial \mathop {\rm{E}}\limits_{{z_1},{z_3}} \left[ {{\varepsilon _{{\rm{MRC}}}}} \right]}{{\partial r}} \\ 
 & + \frac{{\partial ^2 {\mathop {\rm{E}}\limits_{{z_1},{z_3}} \left[ {{\varepsilon _{{\rm{MRC}}}}} \right]}}}{{\partial ^2 r}} \left( {1 -{\mathop {\rm{E}}\limits_{{z_2}} \left[ {{\varepsilon _2}} \right]}} \right). \\ 
 \end{split}
\end{equation}
Hence, $\frac{{{\partial ^2}{C_{{\rm{BL}}}}}}{{{\partial ^2}r}}<0$ if $ \frac{{\partial ^2 {\mathop {\rm{E}}\limits_i \left[ {{\varepsilon _{{\rm{R}},i}}} \right]}}}{{\partial ^2 r}}>0$.

Recall that the coding rate is determined based on $ \eta {{\bar h}}_{j}^2$ ($j=1,2,3$), where  $0 < \eta  \le  \hat z$. 
In other words, the coding rate is definitely lower than the Shannon capacity of a link (either the backhaul link or the combined link) if the fading gain of this link is higher than $\eta$. 
Consider for example  the backhaul link with fading $z_2$.  We have $r<{\mathop{\rm C}\nolimits} ({z_2}{|\bar h_2|^2}), z_2 \in (\eta, + \infty)$ and  $r>{\mathop{\rm C}\nolimits} ({z_2}{|\bar h_2|^2}), z_2 \in (0, \eta)$. 
Denote the integral part in~\eqref{eq:second_Derivative_backhaul} as $\Delta \left( {{z_2}} \right)$ for short.  Hence we have $\frac{{{\partial ^2}\mathop {\rm{E}}\limits_{{z_2}} \left[ {{\varepsilon _2}} \right]}}{{{\partial ^2}r}} = \int\nolimits_0^\eta  {\Delta \left( {{z_2}} \right)} d{z_2} + \int\nolimits_\eta ^{ \hat z} {\Delta \left( {{z_2}} \right)} d{z_2} + \int\nolimits_{\hat z}^{ + \infty } {\Delta \left( {{z_2}} \right)} d{z_2}$ where $\int\nolimits_0^\eta  {\Delta \left( {{z_2}} \right)} d{z_2} <0$ $ \int\nolimits_\eta ^{ \hat z} {\Delta \left( {{z_2}} \right)} d{z_2} >0$ and $\int\nolimits_{\hat z}^{ + \infty } {\Delta \left( {{z_2}} \right)} d{z_2}>0$.

As median, $\hat z$ satisfies $\int\nolimits_0^{\hat z} {e^{ - {{z_2 }}} dz_2 }  = \int\nolimits_{\hat z }^{ + \infty } {e^{ - {{z_2 }}} dz_2 }  = \frac{1}{2}$.

The error probability $\varepsilon_2$ satisfies $0 \le {\varepsilon _2}\left( {{z_2}} \right) \le 1$, and in particular $0 \le {\varepsilon _2}\left( {{z_2}} \right) \le \frac{1}{2}$ if the instantaneous channel gain~$z_2$ is higher than the weighted average channel gain (based on which the source determines the coding rate), i.e., ${z_2} > \hat z_2~\ge~\eta$.

$\Rightarrow$  $\mathop {\rm{E}}\limits_{{z_2}} \left[ {{\varepsilon _2}} \right] = \int_0^{ \hat z } {{e^{ - {z_2}}}} {\varepsilon _2}\left( {{z_2}} \right)d{z_2} + \int_{\hat z}^\infty  {{e^{ - {z_2}}}} {\varepsilon _2}\left( {{z_2}} \right)d{z_2}
 < \int_0^{\hat z} {{e^{ - {z_2}}}} d{z_2} + \int_{\hat z}^\infty  {{e^{ - {z_2}}}} \frac{1}{2}d{z_2} = \frac{1}{2}+ \frac{1}{4}=\frac{3}{4}$.

In addition, $\frac{{\partial ^2{\mathop {\rm{E}}\limits_{{z_2}} \left[ {{\varepsilon _2}} \right]}}}{{\partial ^2 r}}$ can be given as:
\begin{equation} 
\label{eq:second_Derivative_backhaul}
\frac{{\partial ^2 {\mathop {\rm{E}}\limits_{{z_2}} \left[ {{\varepsilon _2}} \right]}}}{{\partial ^2 r}} = \mathop {\rm{E}}\limits_{{z_2}} \left[ {\frac{{{\partial ^2}{\varepsilon _2}}}{{{\partial ^2}r}}} \right] =\!\! \int\limits_0^{ + \infty } \!\! {\frac{{{m^{\frac{3}{2}}}\left( {{\rm{C}}({z_2}|\bar h_2|^2) - r} \right){e^{ - \frac{{m{{\left( {{\rm{C}}({z_2}|\bar h_2|^2) - r} \right)}^2}}}{{2\left( {1 - {2^{ - 2{\rm{C}}({z_2}|\bar h_2|^2)}}} \right){{\left( {{{\log }_2}e} \right)}^2}}}}}}}{{\sqrt {2\pi } {{\left( {1 - {2^{ - 2{\rm{C}}({z_2}|\bar h_2|^2)}}} \right)}^{\frac{3}{2}}}{{\left( {{{\log }_2}e} \right)}^3}}}} {e^{ - {z_2}}}d{z_2}. 
\end{equation}

Moreover, the following relationship holds based on $\hat z$:

$\begin{array}{r}
 \int\limits_{\hat z}^{ + \infty } {\frac{{\left( {{\mathop{\rm C}\nolimits}({z_2}{|\bar h_2|^2}) - r} \right)e^{ - \frac{{\left( {{\mathop{\rm C}\nolimits}({z_2}{|\bar h_2|^2}) - r} \right)^2 }}{{\left( {1 - 2^{ - 2{\mathop{\rm C}\nolimits}({z_2}{|\bar h_2|^2})} } \right)}}} }}{{\left( {1 - 2^{ - 2{\mathop{\rm C}\nolimits}({z_2}{|\bar h_2|^2})} } \right)^{\frac{3}{2}} }}} e^{ - {{z_2 }}} dz_2    > \int\limits_0^{\hat z} {\frac{{\left| {{\mathop{\rm C}\nolimits}({z_2}{|\bar h_2|^2}) - r} \right|e^{ - \frac{{\left( {{\mathop{\rm C}\nolimits}({z_2}{|\bar h_2|^2}) - r} \right)^2 }}{{\left( {1 - 2^{ - 2{\mathop{\rm C}\nolimits}({z_2}{|\bar h_2|^2})} } \right)}}} }}{{\left( {1 - 2^{ - 2{\mathop{\rm C}\nolimits}({z_2}{|\bar h_2|^2})} } \right)^{\frac{3}{2}} }}} e^{ - {{z_2 }}} dz_2  \\ 
  \ge \int\limits_0^{\eta  } {\frac{{\left| {{\mathop{\rm C}\nolimits}({z_2}{|\bar h_2|^2}) - r} \right|e^{ - \frac{{\left( {{\mathop{\rm C}\nolimits}({z_2}{|\bar h_2|^2}) - r} \right)^2 }}{{\left( {1 - 2^{ - 2{\mathop{\rm C}\nolimits}({z_2}{|\bar h_2|^2})} } \right)}}} }}{{\left( {1 - 2^{ -  2{\mathop{\rm C}\nolimits}({z_2}{|\bar h_2|^2})} } \right)^{\frac{3}{2}} }}} e^{ - {{z_2 }}} dz_2 >0  \\ 
\end{array}$

$\Rightarrow$  $\int\nolimits_{\hat z}^{ + \infty } {\Delta \left( {{z_2}} \right)} d{z_2} - \left|\int\nolimits_0^\eta  {\Delta \left( {{z_2}} \right)} d{z_2} \right| >0 $.

$\Rightarrow$   We have $\frac{{{\partial ^2}\mathop {\rm{E}}\limits_{{z_2}} \left[ {{\varepsilon _2}} \right]}}{{{\partial ^2}r}}>0$ for the backhaul link.

Regarding the combined link,  the combined channel gain is $|{\bar h_{{\rm{MRC}}}}{|^2}{\rm{ = }}{z_1}|{\bar h_1}{|^2} + {z_3}|{\bar h_3}{|^2}$. Therefore, we have:

$ \begin{array}{l}
 \int\limits_{\hat z }^{  +  \infty } {\int\limits_{\hat z }^{  +  \infty }  {\frac{{\sqrt {2\pi } \left( {{\rm{C}}\left( {|{{\bar h}_{{\rm{MRC}}}}{|^2}} \right) - r} \right)e\!\!^{  \!- \! \frac{{m\left( {{\rm{C}}\left( {|{{\bar h}_{{\rm{MRC}}}}{|^2}} \right) - r} \right)^2 }}{{\left( {1 - 2^{ - 2{\rm{C}}\left( {|{{\bar h}_{{\rm{MRC}}}}{|^2}} \right)} } \right)\left( {\log _2 e} \right)^2 }}  \!- \! {{z_{1} }}  \!- \! {{z_{3} \! }}} }}{{\left( {\frac{{1  \!- \! 2^{  \!- \! 2{\rm{C}}\left( {|{{\bar h}_{{\rm{MRC}}}}{|^2}} \right)} }}{m}} \right)^{\frac{3}{2}} \left( {\log _2 e} \right)^3 }}} dz_{1} dz_{3} }  \\ 
  \ge  \int\limits_0^{\eta } {\int\limits_0^{\eta }   {\frac{{\sqrt {2\pi } \left| {{\rm{C}}\left( {|{{\bar h}_{{\rm{MRC}}}}{|^2}} \right)  - r} \right|e \! \!^{ - \frac{{m\left( {{\rm{C}}\left( {|{{\bar h}_{{\rm{MRC}}}}{|^2}} \right)  \!- \! r} \right)^2 }}{{2\left( {1 - 2^{ - 2{\rm{C}}\left( {|{{\bar h}_{{\rm{MRC}}}}{|^2}} \right)} } \right)\left( {\log _2 e} \right)^2 }}  \!- \! {{z_{1} }}  \!- \! {{z_{3} \! }}} }}{{4\left( {\frac{{1 - 2^{  \!- \! 2{\rm{C}}\left( {|{{\bar h}_{{\rm{MRC}}}}{|^2}} \right)} }}{m}} \right)^{\frac{3}{2}} \left( {\log_2 e} \right)^3 }}} dz_{1} dz_{3} }   > 0  
 \end{array}$

Similarly, we have  $\frac{{{\partial ^2}\mathop {\rm{E}}\limits_{{z_1},{z_3}} \left[ {{\varepsilon _{{\rm{MRC}}}}} \right]}}{{{\partial ^2}r}} > 0$.




%

Then, to prove $ \frac{{\partial ^2 {\mathop {\rm{E}}\limits_i \left[ {{\varepsilon _{{\rm{R}},i}}} \right]}}}{{\partial ^2 r}}>0$, the following  two cases are considered, which differ in the relationship between the average channel gains of the backhaul link and the combined link are considered. 
The first case is $|\bar h_2|^2\le |\bar h_1|^2  +|\bar h_3|^2$,  where the average channel gain of the backhaul link is lower than the combined link.
As the fading of different links are i.i.d., under this case the instantaneous channel gain of the backhaul link is more likely to be lower than the combined link.


$|\bar h_2|^2 \le |\bar h_1|^2  +|\bar h_3|^2$

$\Rightarrow$ $3/4>\mathop {\rm{E}}\limits_{{z_2}} \left[ {{\varepsilon _2}} \right] \ge \mathop {\rm{E}}\limits_{{z_1},{z_3}} \left[ {{\varepsilon _{{\rm{MRC}}}}} \right]$ and it is easy to prove $\frac{{{\partial ^i}\mathop {\rm{E}}\limits_{{z_2}} \left[ {{\varepsilon _2}} \right]}}{{{\partial ^i}r}} \ge \frac{{{\partial ^i}\mathop {\rm{E}}\limits_{{z_1},{z_3}} \left[ {{\varepsilon _{{\rm{MRC}}}}} \right]}}{{{\partial ^i}r}},i = 1,2,... + \infty $ based  on~\eqref{eq:single_link_error_pro},~\eqref{eq:expected_error_single_2} and~\eqref{eq:expected_error_combine_link}.

$\Rightarrow$ Based on~\eqref{eq:second_de_overall_error_pro}, $\frac{{\partial ^2 {\mathop {\rm{E}}\limits_i \left[ {{\varepsilon _{{\rm{R}},i}}} \right]}}}{{\partial ^2 r}}$ is bounded by:

$\frac{{{\partial ^2}\mathop {\rm{E}}\limits_i \left[ {{\varepsilon _{{\rm{R}},i}}} \right]}}{{{\partial ^2}r}}{{   > }} \frac{1}{4}\frac{{{\partial ^2}\mathop {\rm{E}}\limits_{{z_2}} \left[ {{\varepsilon _2}} \right]}}{{{\partial ^2}r}} - 2{\left( {\frac{{\partial \mathop {\rm{E}}\limits_{{z_2}} \left[ {{\varepsilon _2}} \right]}}{{\partial r}}} \right)^2} $.




Consider $m$ is the blocklength, where $m \gg 1$, we have:

 $ \frac{{{\!\partial ^2}\!\!\mathop {\rm{E}}\limits_i \left[ {{\varepsilon _{{\rm{R}},i}}} \right]}}{{{\partial ^2}r}} \!\! >\!\! \frac{{{m^{\frac{3}{2}}}}}{{8\!\sqrt {2\pi } {{\left( {{{\log }_2}e} \right)}^{\!3\!}}}}\!\!\int\limits_0^{  +  \infty } \!\!{\frac{{\left( {{\rm{C}}(\!{z_2}|\bar h_2|^2)  -  r} \right){e\!\!^{  -  \frac{{m{{\left( {{\rm{C}}(\!{z_2}|\bar h_2|^2)  -  r} \right)}^2  \!\!}}}{{2\left( {1  -  {2^{  -  2{\rm{C}}(\!{z_2}|\bar h_2|^2)}}} \right){{\left( {{{\log }_2}e} \right)}^2 }}}  \!\!}}}}{{{{\left( {1  -  {2^{ - 2{\rm{C}}({z_2}|\bar h_2|^2)}}} \right)}^{\frac{3}{2}}}}}{e^{  -  {z_2}}}} d{z_2}$ .
\rightline{$\quad \quad \quad \quad \quad \quad\quad \quad\quad=\!\frac{{{m^{\frac{3}{2}}}}}{{8\sqrt {2\pi } {{\left( {{{\log }_2}e} \right)}^3}}}\int\limits_0^{ + \infty } {\Phi \left( \!{{z_2}} \right){e^{\! - \!{z_2}}}} d{z_2}$}.


Based on the same idea during the above proof of $\frac{{{\partial ^2}\mathop {\rm{E}}\limits_{{z_2}} \left[ {{\varepsilon _2}} \right]}}{{{\partial ^2}r}}>0$, it is easy to have $\int\nolimits_{\hat z}^{ + \infty } {\Phi \left( {{z_2}} \right)} d{z_2} - \left|\int\nolimits_0^\eta  {\Phi \left( {{z_2}} \right)} d{z_2} \right| >0 $ and $\int\nolimits_\eta^{\hat z}   \Phi \left( {{z_2}} \right)d{z_2}>0$.

$\Rightarrow$ $\int\nolimits_0^\infty   \Phi \left( {{z_2}} \right)d{z_2}>0$.

$\Rightarrow$  $\frac{{\partial ^2 {\mathop {\rm{E}}\limits_i \left[ {{\varepsilon _{{\rm{R}},i}}} \right]}}}{{\partial ^2 r}}> 0$  under the case $|\bar h_2|^2\le |\bar h_1|^2  +|\bar h_3|^2$.

Under the other case $|\bar h_2|^2> |\bar h_1|^2  +|\bar h_3|^2$, it can be proved $\frac{{\partial ^2 {\mathop {\rm{E}}\limits_i \left[ {{\varepsilon _{{\rm{R}},i}}} \right]}}}{{\partial ^2 r}}> 0$ similarly (based on $\frac{{{\partial ^2}\mathop {\rm{E}}\limits_{{z_1},{z_3}} \left[ {{\varepsilon _{{\rm{MRC}}}}} \right]}}{{{\partial ^2}r}} >~0$).

$\Rightarrow$  $\frac{{{\partial ^2}{C_{{\rm{BL}}}}}}{{{\partial ^2}r}} <0$.

$\Rightarrow$ ${C_{\rm{BL}}}$ is a strictly concave in the coding rate.

\end{proof}

\section{Proof of the Corollary 1}

\begin{proof}
  $r$ is strictly increasing in $\eta$, $0 < \eta  \le \hat z$.
\\
$\Rightarrow$ $\forall$  $x <y$,  $x, y \in (0,\hat z)$ and $  \lambda   \in [0,1]$, we have ${\left. r \right|_{\eta = x}} < {\left. r \right|_{\eta = \lambda x + \left( {1 - \lambda } \right)y}} < {\left. r \right|_{\eta = x}}$.
\\
 Based on Theorem  1, ${C_{{\text{BL}}}}$ is concave in $r$,
\\
$\Rightarrow$ $\min \left\{ {{C_{{\text{BL}}}}\left( {{{\left. r \right|}_{\eta = x}}} \right),{C_{{\text{BL}}}}\left( {{{\left. r \right|}_{\eta = y}}} \right)} \right\} \leqslant {C_{{\text{BL}}}}\left( {{{\left. r \right|}_{\eta = \lambda x + \left( {1 - \lambda } \right)y}}} \right)$. 
\\
$\Rightarrow$  ${C_{{\text{BL}}}}$ is quasi-concave in $\eta$,  $0 < \eta  \le \hat z$.

\end{proof}

\section{Proof of the Theorem  2}

\begin{proof}
Comparing~\eqref{eq:MSDR_final} and~\eqref{eq: BL-throughput_over_fading}, we have:

$ {{R_{{\rm{MS}}}}}  \!=\! \frac{{{C_{{\rm{BL}}}}}}{2}  +  {R^*}\!$, 
where $\!{R^*} \!\!=\! \frac{r}{4}\!\sqrt {{{\!( {1\! - \! \mathop {\rm{E}}\limits_{{z_1}\!,{z_2}\!,{z_3}\!}  \left[ {{\varepsilon _{\rm{R}}}} \right] } )}^{\!2\!}}  + \! \frac{{4m\ln \left(\! {{P_{\rm{d}}}} \!\right)}}{d}   \mathop {\rm{E}}\limits_{{z_1}\!,{z_2}\!,{z_3}\!} \!\! \!\left[ {{\varepsilon _{\rm{R}}}} \right]  \!(\!1 \! -  \mathop {\rm{E}}\limits_{{z_1}\!,{z_2}\!,{z_3}\!}   \left[ {{\varepsilon _{\rm{R}}}} \right] )} $.

As we have proved that $C_{{\rm{BL}}}$ is concave in $r$ in Theorem  1, Theorem  2 holds if ${R^*}$ is concave, too.   

Note that $\frac{{4m\ln \left( {{P_{\rm{d}}}} \right)}}{d}$ is not influenced by $r$. It is actually a negative constant as $\ln \left( {{P_{\rm{d}}}} \right) <0$. 
We denote this constant by $\varphi $ $ (\varphi < 0)$ for short. To facilitate the proof we also denote $ \mathop {\rm{E}}\limits_{{z_1},{z_2},{z_3}} \left[ {{\varepsilon _{\rm{R}}}} \right] $ by  ${\overline \varepsilon  _{\rm{R}}}$  for short.

Then, we have: 
$\!{R^*}  = \frac{r}{4}\!\sqrt {{{( \!{1  -  {\overline \varepsilon  _{\rm{R}}}} )}^{\!2\!}} \! +  \varphi {\overline \varepsilon  _{\rm{R}}}(1  -  {\overline \varepsilon  _{\rm{R}}})}  \!=\! \frac{r}{4}\!\sqrt {\!1  + \! \left( {\varphi   -  2} \right)\!{\overline \varepsilon  _{\rm{R}}}  +  \left( {1  -  \varphi } \right){{\overline \varepsilon  _{\rm{R}}}}^{\!2\!}}  $

As $R^*$ is a square root function, it should be satisfied that:
\begin{equation} 
\label{eq:square_root_inequality}
 {{{( {1 - {\overline \varepsilon  _{\rm{R}}}} )}^2} + \varphi  {\overline \varepsilon  _{\rm{R}}} \cdot (1 - {\overline \varepsilon  _{\rm{R}}})} \ge 0.
\end{equation} 

 ${\overline \varepsilon  _{\rm{R}}}$ is the expectation of the error probability over fading

 $\Rightarrow$  $0 < {\overline \varepsilon  _{\rm{R}}} < 1$

 $\Rightarrow$  $1- {\overline \varepsilon  _{\rm{R}}} > 0$. Combining this with~\eqref{eq:square_root_inequality}

 $\Rightarrow$ $1 - {\overline \varepsilon  _{\rm{R}}} + \varphi {\overline \varepsilon  _{\rm{R}}} \ge 0$,

 $\Rightarrow$ $1 \ge \left( 1 -\varphi \right){\overline \varepsilon  _{\rm{R}}}$.

$\frac{{\partial {R^*}}}{{\partial r}} = \frac{1}{4}\sqrt {1 + \left( {\varphi  - 2} \right){\overline \varepsilon  _{\rm{R}}} + \left( {1 - \varphi } \right){{\overline \varepsilon  _{\rm{R}}}}^2}  + \frac{r}{8}\frac{{\left( {\varphi  - 2} \right) + 2\left( {1 - \varphi } \right){\overline \varepsilon  _{\rm{R}}}}}{{\sqrt {1 + \left( {\varphi  - 2} \right){\overline \varepsilon  _{\rm{R}}} + \left( {1 - \varphi } \right){{\overline \varepsilon  _{\rm{R}}}}^2} }}\frac{{\partial {\overline \varepsilon  _{\rm{R}}}}}{{\partial r}}$,

$ \frac{{{\partial ^2}{R^*}}}{{{\partial ^2}r}} = \frac{1}{8}\frac{{\left( {\varphi  - 2} \right) + 2\left( {1 - \varphi } \right){{\bar \varepsilon }_{\rm{R}}} + r\left( {1 - \varphi } \right)}}{{{{\left( {{{\left( {1 - {{\bar \varepsilon }_{\rm{R}}}} \right)}^2} + \varphi {{\bar \varepsilon }_{\rm{R}}}(1 - {{\bar \varepsilon }_{\rm{R}}})} \right)}^{\frac{1}{2}}}}}
  +  \frac{r}{{16}}\frac{{  -  {{\left( {\left( {\varphi   -  2} \right)  +  2\left( {1  -  \varphi } \right){{\bar \varepsilon }_{\rm{R}}}} \right)}^2}}}{{{{\left( {{{\left( {1  -  {{\bar \varepsilon }_{\rm{R}}}} \right)}^2}  +  \varphi {{\bar \varepsilon }_{\rm{R}}}(1  -  {{\bar \varepsilon }_{\rm{R}}})} \right)}^{\frac{3}{2}}}}}{\left( {\frac{{\partial {{\bar \varepsilon }_{\rm{R}}}}}{{\partial r}}} \right)^2} \! +  \!\frac{r}{8}\frac{{\left( {\varphi   -  2} \right)  +  2\left( {1  -  \varphi } \right){{\bar \varepsilon }_{\rm{R}}}}}{{{{\left( {{{\left( {1  -  {{\bar \varepsilon }_{\rm{R}}}} \right)}^2}  +  \varphi {{\bar \varepsilon }_{\rm{R}}}(1  -  {{\bar \varepsilon }_{\rm{R}}})} \right)}^{\frac{1}{2}}}}}\frac{{{\partial ^2}{{\bar \varepsilon }_{\rm{R}}}}}{{{\partial ^2}r}}$.

As we have $\varphi  < 0$ and  $0 < {\overline \varepsilon  _{\rm{R}}} < 1$

 $\Rightarrow$  $\!0<{( {1 - {\overline \varepsilon  _{\rm{R}}}} )^2} + \varphi {\overline \varepsilon  _{\rm{R}}}(1 - {\overline \varepsilon  _{\rm{R}}}) < {( {1 - {\overline \varepsilon  _{\rm{R}}}} )^2} < 1$

 $\Rightarrow$  
$\!0\!<\!{{{\left( {{{(\! {1\! -\! {\overline \varepsilon  _{\rm{R}}}} )}^{2}} \! + \! \varphi {\overline \varepsilon  _{\rm{R}}}(1 \! -  {\overline \varepsilon  _{\rm{R}}})} \right)}^{!\frac{3}{2}}}}< {{{\left( {{{( {1  -  {\overline \varepsilon  _{\rm{R}}}} )}^{2\!}}  +  \varphi {\overline \varepsilon  _{\rm{R}}}(1\ -\! {\overline \varepsilon  _{\rm{R}}})} \right)}^{\frac{1}{2}}}\!}<1      \!\!$

 $\Rightarrow$  we have:

$\frac{{{\partial ^2}{R^*}}}{{{\partial ^2}r}} < \frac{1}{8}\frac{{\left( {\varphi  - 2} \right) + 2\left( {1 - \varphi } \right){{\bar \varepsilon }_{\rm{R}}} + r\left( {1 - \varphi } \right)}}{{{{\left( {{{\left( {1 - {{\bar \varepsilon }_{\rm{R}}}} \right)}^2} + \varphi {{\bar \varepsilon }_{\rm{R}}}(1 - {{\bar \varepsilon }_{\rm{R}}})} \right)}^{\frac{1}{2}}}}} 
  +  \frac{r}{{16}}\frac{{-{{\left( {\left( {\varphi   -  2} \right)  +  2\left( {1 -\varphi } \right){{\bar \varepsilon }_{\rm{R}}}} \right)}^2}}}{{{{\left( {{{\left( {1  -  {{\bar \varepsilon }_{\rm{R}}}} \right)}^2}  +  \varphi {{\bar \varepsilon }_{\rm{R}}}(1 - {{\bar \varepsilon }_{\rm{R}}})} \right)}^{\frac{1}{2}}}}}{\left( {\frac{{\partial {{\bar \varepsilon }_{\rm{R}}}}}{{\partial r}}} \right)^2} \! + \! \frac{r}{8}\frac{{\left( {\varphi   -  2} \right)  +  2\left( {1  -  \varphi } \right){{\bar \varepsilon }_{\rm{R}}}}}{{{{\left( {{{\left( {1  -  {{\bar \varepsilon }_{\rm{R}}}} \right)}^2}  +  \varphi {{\bar \varepsilon }_{\rm{R}}}(1  -  {{\bar \varepsilon }_{\rm{R}}})} \right)}^{\frac{1}{2}}}}}\frac{{{\partial ^2}{{\bar \varepsilon }_{\rm{R}}}}}{{{\partial ^2}r}}$.

As $1 \ge \left( 1 -\varphi \right){\overline \varepsilon  _{\rm{R}}}$, we have:

${\frac{{{\partial ^2}{R^*}}}{{{\partial ^2}r}} <\frac{{2\varphi  + 2r\left( {1 -\varphi - \frac{{{\varphi ^2}}}{2}{{\left( {\frac{{\partial {{\bar \varepsilon }_{\rm{R}}}}}{{\partial r}}} \right)}^2} + \varphi \frac{{{\partial ^2}{{\bar \varepsilon }_{\rm{R}}}}}{{{\partial ^2}r}}} \right)}}{{16{{\left( {{{\left( {1 - {{\bar \varepsilon }_{\rm{R}}}} \right)}^2} + \varphi {{\bar \varepsilon }_{\rm{R}}}(1 - {{\bar \varepsilon }_{\rm{R}}})} \right)}^{\frac{1}{2}}}}}}$.

As shown in Appendix A,  $\frac{{\partial ^2 {\mathop {\rm{E}}\limits_i \left[ {{\varepsilon _{{\rm{R}},i}}} \right]}}}{{\partial ^2 r}}>0$. In addition,  ${\left( {\frac{{\partial {{\bar \varepsilon }_{\rm{R}}}}}{{\partial r}}} \right)^2} \sim {\rm O}\left( m \right)$ and $\frac{{{\partial ^2}{{\bar \varepsilon }_{\rm{R}}}}}{{{\partial ^2}r}} \sim {\rm O}\left( {{m^{3/2}}} \right)$. 
Moreover, $\varphi $ is a constant with reasonable value\footnote{Recall that $\varphi$ is subject to the QoS constraints  $\{$delay (in symbols), delay violation probability$\}$. Considering  extremely loose constraints $\{100m, 10^{-0.5} \}$ and extremely strict  constraints $\{2m, 10^{-7} \}$ (recall that the length of a transmission period of relaying is $2m$), we have $\varphi \in (-30m,-0.05m)$.}  $\varphi \in (-30m,-0.05m)$.
Therefore,  $\frac{{{\partial ^2}{R^*}}}{{{\partial ^2}r}} <0$ holds. 

Hence,  ${{R_{{\rm{MS}}}}} $ is concave in $r$.

\end{proof}

\



\bibliographystyle{IEEEtran}


\end{document}